\documentclass[preprint,12pt,authoryear]{elsarticle}

\usepackage{amssymb}
\usepackage{amsthm}
\usepackage{amsmath}
\usepackage{rotating}

\usepackage[T1]{fontenc}
\usepackage{multirow}
\usepackage{subfig}

\begin{document}

\begin{frontmatter}


\title{A method for supervoxel-wise association studies of age and other non-imaging variables from coronary computed tomography angiograms}



\author[aff1]{Johan \"{O}fverstedt}
\author[aff1]{Elin Lundstr\"{o}m}
\author[aff3,aff4]{G\"{o}ran Bergstr\"{o}m}
\author[aff1,aff2]{Joel Kullberg\corref{cor1}}
\author[aff1,aff2]{H{\aa}kan Ahlstr\"{o}m\corref{cor1}}
\cortext[cor1]{These authors contributed equally to this work and share senior authorship.}

\affiliation[aff1]{organization={Radiology, Department of Surgical Sciences, Uppsala University}, city={Uppsala}, country={Sweden}}
\affiliation[aff2]{organization={Antaros Medical}, city={M\"{o}lndal}, country={Sweden}}
\affiliation[aff3]{organization={Department of Molecular and Clinical Medicine, Institute of Medicine, Sahlgrenska Academy, University of Gothenburg}, city={Gothenburg}, country={Sweden}}
\affiliation[aff4]{organization={Department of Clinical Physiology, Sahlgrenska University Hospital, Region V{\"a}stra G{\"o}taland}, city={Gothenburg}, country={Sweden}}

\begin{abstract}
The study of associations between an individual's age and imaging and non-imaging data is an active research area that attempts to aid understanding of the effects and patterns of aging. In this work we have conducted a supervoxel-wise association study between both volumetric and tissue density features in coronary computed tomography angiograms and the chronological age of a subject, to understand the localized changes in morphology and tissue density with age. To enable a supervoxel-wise study of volume and tissue density, we developed a novel method based on image segmentation, inter-subject image registration, and robust supervoxel-based correlation analysis, to achieve a statistical association study between the images and age. We evaluate the registration methodology in terms of the Dice coefficient for the heart chambers and myocardium, and the inverse consistency of the transformations, showing that the method works well in most cases with high overlap and inverse consistency. In a sex-stratified study conducted on a subset of $n=1388$ images from the SCAPIS study, the supervoxel-wise analysis was able to find localized associations with age outside of the commonly segmented and analyzed sub-regions, and several substantial differences between the sexes in the association of age and volume.
\end{abstract}







\end{frontmatter}


\newcommand{\abs}[1]{|#1|}

\section{Introduction}
\label{sec:introduction}

Cardiovascular diseases (CVDs) are among the leading causes of death worldwide. The development of biomarkers that are well associated with disease and abnormalities is essential to improve prevention, therapies, and the outcomes of patients. Imaging is used in a multitude of ways in research and clinical practices related to CVD, such as finding associations between image-derived features and CVDs \citep{fagman2023high,chen2023radiomics}, or quantification of clinically relevant measures from image data \citep{agatston1990quantification}. Another approach to understanding the patterns of disease is to target a different major aspect affecting the morphology and function of the heart, the process of aging, with the goal of quantifying features of aging of normal (healthy) and abnormal hearts.

Studying the associations between aging and different imaging (and non-imaging) data sources remains an active research area. One possible method for such studies is to segment sub-regions of a body or organ, measuring the volume and image statistics in each sub-region, and investigating the association of those variables to age \citep{bai2020population,wasserthal2023totalsegmentator}. In  \citep{bai2020population}, associations between chronological age and image-based heart morphology phenotypes derived from over 30000 cardiac magnetic resonance images, are studied. The findings include significant relationships between age and heart features (e.g. left ventricle volume, in agreement with earlier studies \citep{chida1994morphological}, and aorta volume), as well as associations with CVDs. Another approach that has attracted much attention in recent years is the deep learning-based prediction of age from various input modalities, such as MRI \citep{langner2019identifying,ecker2024deep}, and electrocardiography (ECG) \citep{lima2021deep}, and investigating the relevance of this predicted age for health and disease. Each of these approaches has its own set of advantages and disadvantages. Segmenting specific detailed regions requires prior knowledge, high-quality reference regions of interest, or there is a risk that the resulting features are too inconsistent. It may also be limited to regions with expected importance a priori, and not promote discovery of previously unknown regions of importance, the latter being essential for exploratory studies. One one hand, deep learning approaches tend to exhibit good predictive performance \citep{ecker2024deep,kerber2023deep} once trained on sufficient data. On the other hand, they largely work as black-box models which are challenging to derive clear insights from (apart from saliency analysis that points out from where in the input the prediction is derived, but not if the used features where image intensities, texture, shape/volume of structures, etc.). A class of methods that do not rely on specific pre-selected regions of interest, nor on black-box models, is voxel-wise association studies (Imiomics) \citep{strand2017concept}, building on the inter-subject image registration of a cohort of images, to relate the image intensity or local volume of each voxel to a non-imaging variable of interest. Computed Tomography (CT) is a widely used imaging technique in clinical practice and research. It generates images from a spectrum of X-rays where the resulting voxel values are expressed in terms of Hounsfield units (HUs), related to the tissue density. In this work, we develop a methodology to study the aging of the heart using analysis applied to cardiac CT images, with the aim of finding density and morphological features that are associated with the age of the subjects.

\begin{figure}[ht]
    \centering
    \includegraphics[width=1.0\linewidth,trim={1.2cm 3.5cm 1cm 2cm},clip]{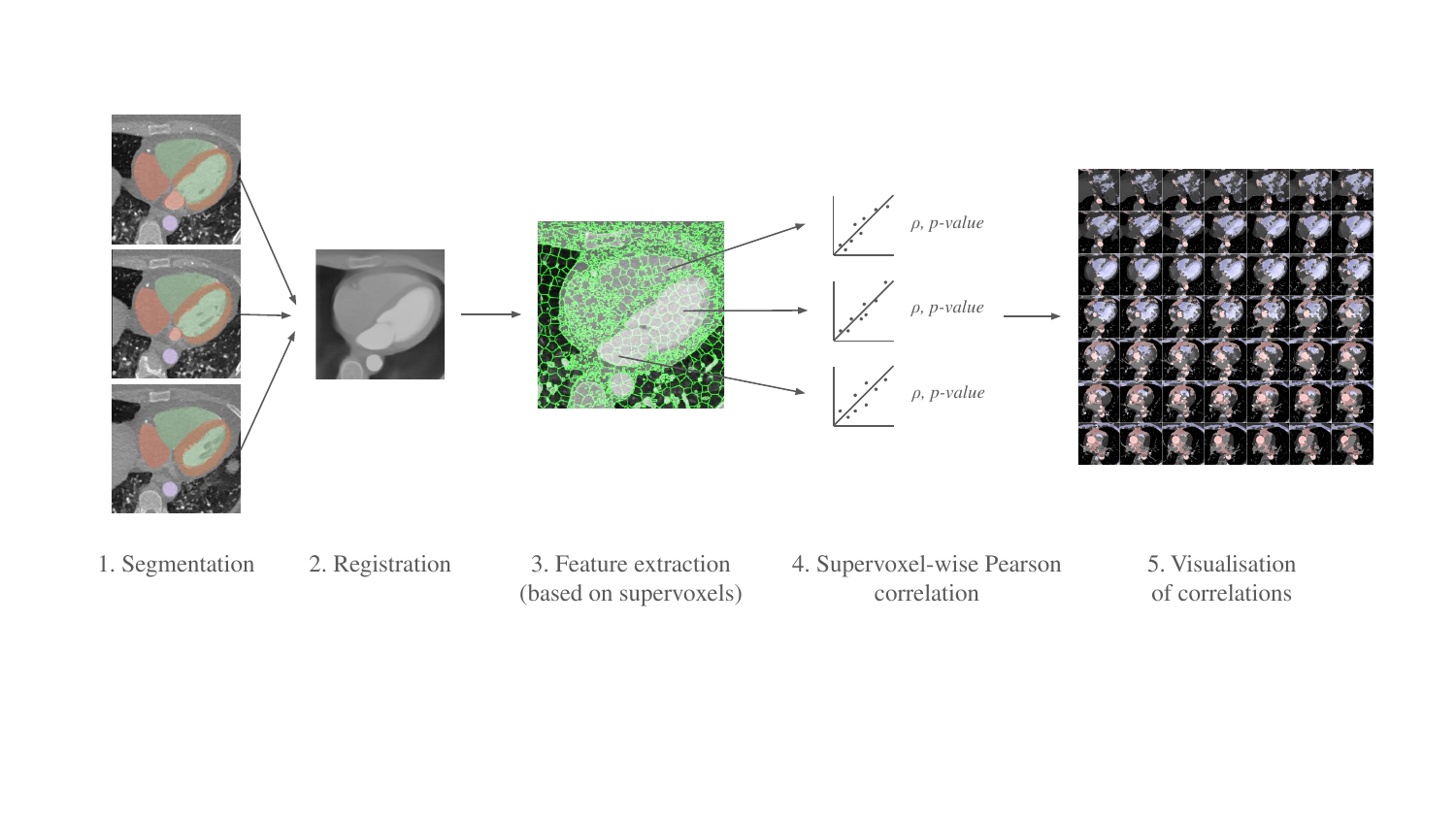}
    \caption{Illustration of the proposed methodology. Step 1: The images are segmented using TotalSegmentator. Step 2: The images are registered to a common reference space using a novel registration method, combining intensity information and segmentation masks. Step 3: Supervoxel-derived aggregate features of local volume and density are extracted for the entire cohort. Step 4: Pearson correlation is performed independently between each supervoxel/feature and a non-imaging variable such as age. Step 5: The statistically significant correlation coefficients are visualized in color overlayed on the original CCTA image volume, providing a detailed map of the association between the variable and sub-regions of the heart.}
    \label{fig:fig1}
\end{figure}

Coronary Computed Tomography Angiography (CCTA) imaging \citep{schmermund2018coronary,antonopoulos2022cardiovascular} is a form of (CT) imaging of the heart using contrast medium enhancement to facilitate the acquisition of detailed (high-resolution) image volumes of the whole heart, chambers, and coronary vessels. In this work, we investigate the possibility of finding associations between chronological age and spatially localized morphological features in these detailed images. CT images have been demonstrated to be well correlated to age \citep{kerber2023deep}, where age estimation from clinical CT images of the thorax (without contrast enhancement or specific focus on the heart) was performed. Voxel-wise integration of CT slices and non-imaging data has been done for body composition analysis \citep{ahmad2024voxelwise}.

We make several technical contributions: (i) We have developed a novel inter-subject deformable image registration method for CCTAs, (ii) we have developed a methodology for supervoxel-wise cohort analysis of CCTAs (density and local volume based on deformable image registration), and (iii) we introduce a more efficient and robust Imiomics-analysis methodology by performing the statistical analysis on cardiac supervoxels (connected clusters of voxels with similar density), rather than on the original voxels. The use of supervoxels speeds up the analysis substantially and increases robustness by reducing the number of statistical tests as well as the noise level of the image values (in terms of imaging noise and alignment error noise) by acting on averages over multiple voxel values rather than on values of individual voxels. An overview of the proposed method is presented in Fig.~\ref{fig:fig1}.

We applied the developed methodology to a subset of subjects from the Swedish Cardiopulmonary Imaging Study (SCAPIS) \citep{bergstrom2015swedish}, a cohort study of Swedish residents. We performed sex-stratified inter-subject registration of the CCTAs to a selected reference image, followed by supervoxel-wise statistical analysis of the density and local volume features and age. As an outcome of the supervoxel-wise analyses conducted in this work, we found spatially varying associations between volume and age, with substantial differences between the patterns of aging between the two sexes. The volume of the left atrium was observed to have a positive significant association with age in females while no significant association was found in males. The volume of the left ventricle had a negative significant association with age in both sexes. Furthermore, we also found a negative association between the density of the fat tissue of the hearts and age in both females and males. By using both explicit measurements via segmentation and supervoxel-wise analysis across the cohort, we found that both methods agree on the most salient associations, increasing the trust in the validity of these associations.

\section{Materials and Methods}

In this work, we used multiple types of image analysis techniques (filtering, segmentation and registration) to study the volume of key regions of the hearts and to obtain a supervoxel-wise cohort analysis of hearts from the SCAPIS study.

\subsection{Data}

SCAPIS is a cohort study of a random sample ($n=30154$) of men and women aged 50-64 years (typically past menopause and before retirement with corresponding changes in lifestyle factors) in Sweden \citep{bergstrom2015swedish}.

For this feasibility study, we used a subset of the full cohort ($n=1388$, $n_{\mathtt{FEMALE}}=722$, $n_{\mathtt{MALE}}=666$) who were examined with CT at Uppsala University Hospital, Uppsala, Sweden. All images were pseudonymized but information on (legal) sex and age in whole months at the time of the examination was maintained.

For each subject, we had access to one 3D CCTA image volume of axial slices (size: $512 \times 512$, number of slices: between 350 and 550 slices), with slice spacing approximately $0.30\texttt{ mm}$, and with a slice thickness typically $0.33\texttt{ mm}$ (with minor variation across the dataset). As part of the CCTA imaging protocol, the subjects were given an intravenous contrast injection before the image acquisition. The time delay between injection and start of the image acquisition was determined from a test bolus to match the start time with the contrast arrival at the ascending aorta, as an attempt to standardize the contrast distribution between subjects.

The median radiation dose for each subject was 4.2 mSv (2 mSv for the CCTA, 2 mSV for the lung examination, and 0.2 mSv for an additional single slice metabolic study) \citep{bergstrom2015swedish}. The kV is either 100 or 120, and mAs is 320 or 340, depending on the sub-protocol used.

\subsection{Ethics}

Ethics approval was obtained from the Swedish Ethical Review Authority (Dnr 2022-07308-01) to conduct this research study related to human subjects, with associated sex and age information. All subjects provided informed written consent for their collected data to be used for research and for that research to be published. The study adheres to the Declaration of Helsinki. SCAPIS has been approved as a multicentre trial by the ethics committee at Umea University and adheres to the Declaration of Helsinki \citep{bergstrom2015swedish}.

\subsection{Cardiac Image Segmentation}

Medical image segmentation is the process of delineating an image into 2 or more regions, e.g., corresponding to selected organs or tissue types. Much has been written about whole-heart segmentation in particular \citep{chen2020deep}, though not many annotated open datasets or already trained segmentation models are published that are readily applicable to new data. Recently, Total Segmentator (TS) \citep{wasserthal2023totalsegmentator} was developed, which is a general, multi-protocol, multi-field-of-view, method (and software) that is capable of segmenting up to 104 different regions of the whole body, including several detailed sub-regions of the heart. For heart segmentation, the authors state that the method performs better on CT heart images with contrast enhancement than without. In this study, we used image segmentation for two main purposes: (i) to directly perform volume and density measurements of key sub-regions of the heart, (ii) to support the image registration and its evaluation.

We applied TS version 1 to segment the SCAPIS CCTA images. The regions that were of primary interest were the left ventricle (LV), left atrium (LA), right ventricle (RV), right atrium (RA), myocardium (MYO), and the aorta (the ascending and descending aortas are combined into a single label by TS). As reported, the Dice Score achieved for LV, LA, RV, RA, MYO, and aorta were 0.955, 0.941, 0.949, 0.939, 0.937, and 0.981, respectively \citep{wasserthal2023totalsegmentator}. In addition, TS generated segmentation masks for the lungs, pulmonary vein, liver, stomach, and esophagus, which were used mostly to aid in the exclusion of non-cardiac structures of the CCTA images.

We used the resulting segmentations to conduct a pairwise correlation analysis relating age to the volumetric and mean density features.

\subsection{Image Registration}

In this work, we used inter-subject CCTA registration \citep{kirisli2010fully} to align a cohort of cardiac images to a selected template image. The main work done on registration of CCTA images have been for the purpose of heart segmentation \citep{kirisli2010fully}. Already having heart segmentations available from TotalSegmentator, with the registration of the images to spatial standardize the cohort being the main goal, we use the segmented structures to simplify the registration process and increase its reliability.

\begin{figure}[ht]
    \centering
    \begin{turn}{90} 
        \;\;\;\;\;\;\;Slice 1       
    \end{turn}
    \subfloat[][CT image]{\includegraphics[width=0.19\textwidth]{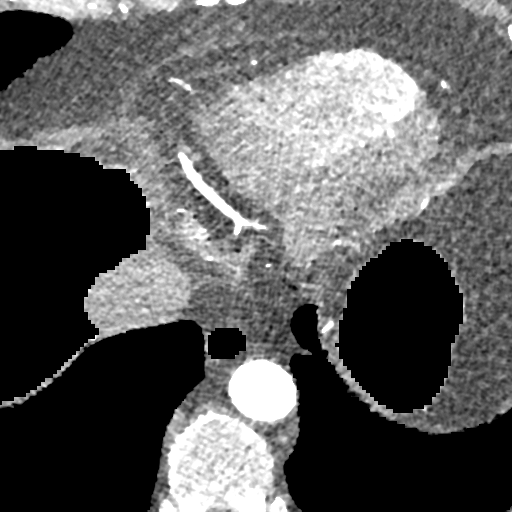}} \hfill
    \subfloat[][Mask: LV, RV, LA, RA]{\includegraphics[width=0.19\textwidth]{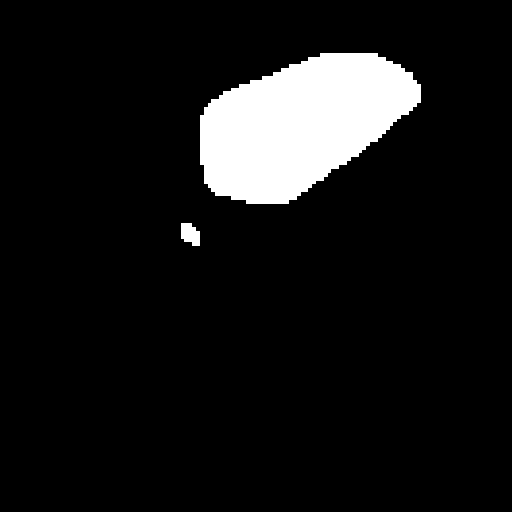}} \hfill
    \subfloat[][Mask: MYO, Aorta]{\includegraphics[width=0.19\textwidth]{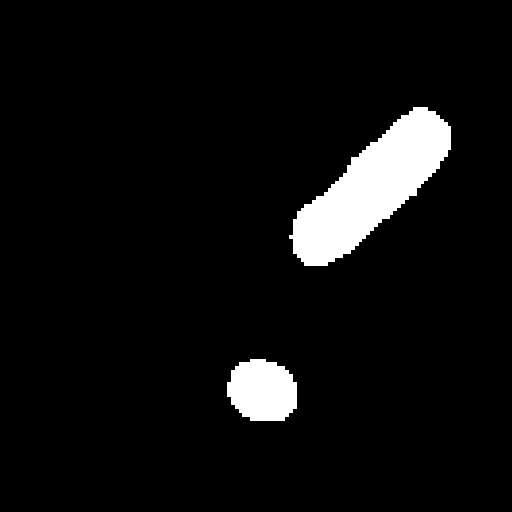}} \hfill
    \subfloat[][Mask: High density tissue]{\includegraphics[width=0.19\textwidth]{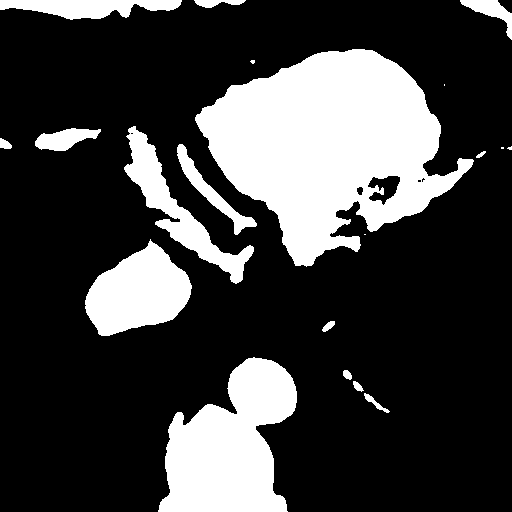}} \hfill
    \subfloat[][Mask: Low density tissue]{\includegraphics[width=0.19\textwidth]{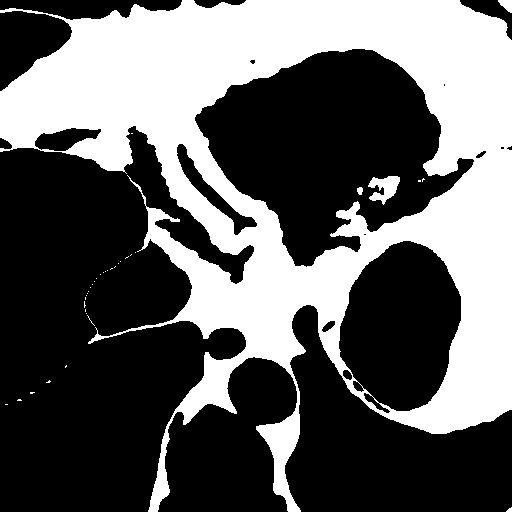}}
    \\
    \begin{turn}{90} 
        \;\;\;\;\;\;\;Slice 2       
    \end{turn}
    \subfloat[][CT image]{\includegraphics[width=0.19\textwidth]{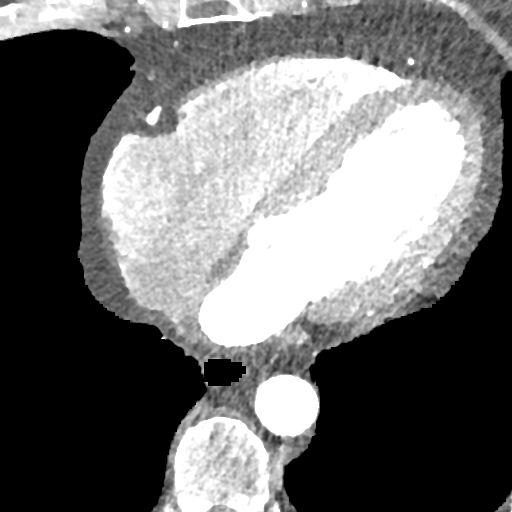}} \hfill
    \subfloat[][Mask: LV, RV, LA, RA]{\includegraphics[width=0.19\textwidth]{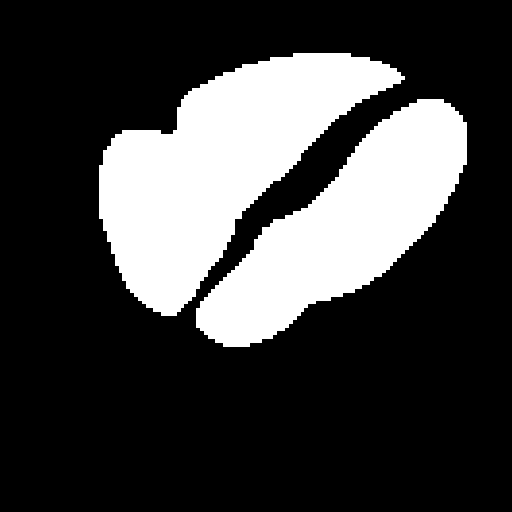}} \hfill
    \subfloat[][Mask: MYO, Aorta]{\includegraphics[width=0.19\textwidth]{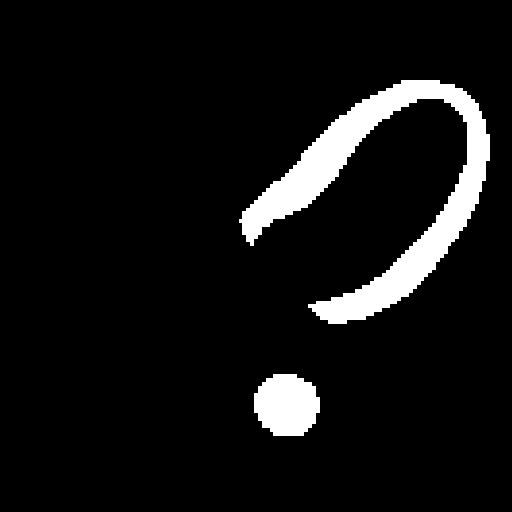}} \hfill
    \subfloat[][Mask: High density tissue]{\includegraphics[width=0.19\textwidth]{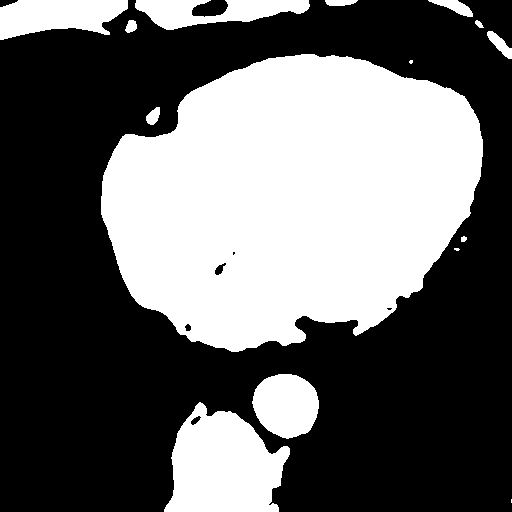}} \hfill
    \subfloat[][Mask: Low density tissue]{\includegraphics[width=0.19\textwidth]{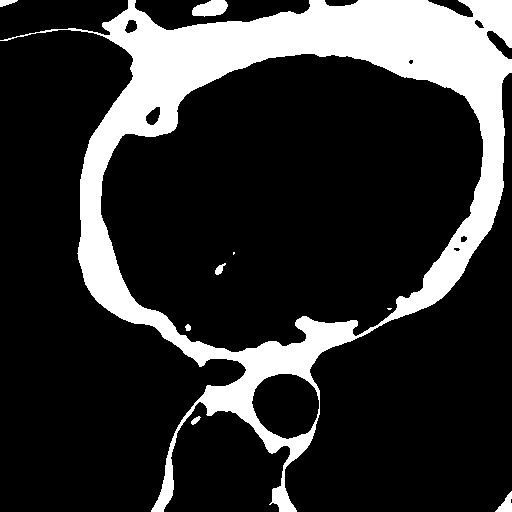}}
    \caption{Illustration of the image channels and masks used in the deformable image registration process for two selected axial slices. (a,f) The pre-processed CT images for two sample axial slices. As can be seen in the black areas, non-cardiac structures such as lungs, stomach, liver, and esophagus have been cut out, and replaced by the minimal density value (after limiting the density range to between -300 and +200 HU). (b,c,g,h) The six anatomical masks used for the image registration (LV, RV, LA, RA, MYO, Aorta), merged into two channels, to reduce the runtime, the amount of memory required, and the number of competing objectives. (d,e,i,j) The low and high-density regions, excluding the non-cardiac structures, as for (a,f), are used to further focus the registration on the relevant cardiac tissue, and aid in the registration of the vessels and other high-density sub-structures.}
    \label{fig:registrationillustration}
\end{figure}

\subsubsection{Affine Transformation Initialization using Bounding Box Matching}

As an initialization for the deformable registration, we computed the axis-aligned bounding boxes (AABB) of the LV mask and computed (in closed form) the scaling and shift required to align them, to obtain the transformation $T_{\texttt{INIT}}$ given by
\begin{gather}
T_{\texttt{INIT}}(x) =
\begin{bmatrix}
    {\frac{e_1^{F}}{e_1^{R}}} & 0 & 0 \\
    0 & {\frac{e_2^{F}}{e_2^{R}}} & 0 \\
    0 & 0 & {\frac{e_3^{F}}{e_3^{R}}} \\
\end{bmatrix}
\begin{bmatrix}
    x_1 \\
    x_2 \\
    x_3 \\
\end{bmatrix}
+
\begin{bmatrix}
    {c_1^{F} - c_1^{R}} \\
    {c_2^{F} - c_2^{R}} \\
    {c_3^{F} - c_3^{R}} \\
\end{bmatrix},
\end{gather}
where $c_d$ denotes the center of mass of the left ventricle bounding box along dimension $d$, and $e_d$ denotes the extent (side-length) of the LV bounding box along dimension $d$. Superscript $R$ denotes the reference image/space and superscript $F$ denotes the floating image/space. 

Given a mostly standardized orientation of the hearts, and at least coarse segmentations of the hearts (in particular without severe errors where false positives occur far from the true extent of the heart), this step ensures that the deformation fields are initialized such that a key structure of the hearts overlap well, up to small rotations and local deformations required to approach a voxel-accurate registration.

\subsubsection{Deformable Image Registration}

Inter-subject deformable image registration  \citep{rueckert2010medical} of CT images is a challenging task due to factors such as differences in anatomy, appearance and contrast, initialization and local optima during optimization \citep{li2006automatic}. 

In this work, we develop a deformable image registration methodology inspired by \citep{jonsson2022image} which incorporates multiple anatomical/tissue-type masks into a multi-objective optimization process, to use additional known semantic information effectively, rather than relying entirely on registration by image (density) values, which can be error-prone due to identical densities from different regions overlapping and forming local minima that are difficult to escape. The registration optimization method used the \emph{deform}\footnote{https://github.com/simonekstrom/deform} framework \citep{ekstrom2020fast,ekstrom2021faster}, and was based on optimization of dense displacement fields by solving minimal graph-cut problems \citep{tang2007non,ekstrom2020fast,ekstromthesis}. The same general methodology (with different semantic masks) has also been applied to inter-subject registration of 2D images from a three-slice CT protocol \citep{ahmad2024voxelwise}, for the purpose of large-scale cohort body composition analysis, where it performed well and enabled detailed voxel-wise analysis.

The CCTA images were pre-processed for the deformable image registration: lungs were removed using the TS lung masks (and the voxels that were within the lung mask were assigned to a density of -1000, which is then clipped to the minimum value in a subsequent pre-processing step) such that the contents within the lungs, consisting of variations of density and numerous vessels would not affect the registration of the heart. The other anatomical regions nearby such as the stomach, liver, and esophagus were similarly removed using the TS masks. Saturation was applied to the images such that all HU below -300 and above +200 were clipped. The HUs were then rescaled by a factor of $\frac{1}{300}$ to assign them to a similar range as the masks.

The deformable image registration method is divided into two main stages. For both stages, we used a combination of the CCTA images and segmentation masks. We combined the main cardiac cavities (LV, RV, LA, RA) into a single mask, and combined the aorta (both ascending and descending) and MYO as a second mask (due to the distance between them with low risk of confusion and the advantage of using a small number of masks). In addition, we used two masks corresponding to groups of tissue classified by the density ($\left[0, \infty\right]$ and $\left[-400, 0\right]$), after pre-processing the input CCTA images with a median filter with radius 4 to reduce their noise-level and avoid many small objects. The pre-processed intensity images and masks are shown in Fig.~\ref{fig:registrationillustration}.

For the first stage, we employed high regularization weights and small block size (which decreases quality but increases speed substantially \citep{ekstrom2020fast} by reducing the size of the sub-regions that each individual graph-cut optimization is performed on), described in detail in Tab.~\ref{tab:deform_param1}, followed by the same method but with low regularization, and slightly different weights assigned to different objectives and larger block size, described in detail in Tab.~\ref{tab:deform_param2}. 

\begin{table}
    \centering
    \caption{Configuration of the first stage of deformable registration.}
    \label{tab:deform_param1}
    \begin{tabular}{l|c}
        Parameter & Value \\
        \hline
        Levels & 6 \\
        Block size & 8 \\
        Regularization Weight & 2.0 \\
        Image Objective Function & NCC (weight 0.25)\\
        Mask Objective Function & SSD \\ 
        Max Iterations & 300, 300, 300, 40, 20, (0) \\
        Mask Weight (LV, RV, LA, RA) & 1.0 \\
        Mask Weight (MYO, Aorta) & 1.0 \\
        Mask Weight (High-density: HU $ \in \left[0, \infty\right]$) & 0.3 \\
        Mask Weight (Low-density: HU $\in \left[-400, 0\right]$) & 0.3 \\
    \end{tabular}
\end{table}

\begin{table}
    \centering
    \caption{Configuration of the second stage of deformable registration.}
    \label{tab:deform_param2}
    \begin{tabular}{l|c}
        Parameter & Value \\
        \hline
        Levels & 6 \\
        Block size & 32 \\
        Regularization Weight & 0.15 \\
        Image Objective Function & NCC (weight 0.5) \\
        Mask Objective Function & SSD \\ 
        Max Iterations & 300, 300, 300, 40, 20, (0) \\
        Mask Weight (LV, RV, LA, RA) & 1.0 \\
        Mask Weight (MYO, Aorta) & 1.0 \\
        Mask Weight (High-density: HU $ \in \left[0, \infty\right]$) & 0.1 \\
        Mask Weight (Low-density: HU $\in \left[-400, 0\right]$) & 0.1 \\
    \end{tabular}
\end{table}

\subsubsection{Template Selection}

An important first step before image registration of entire cohorts to a common space, in the absence of a completely unbiased and representative atlas, is to select a template subject to act as a reference. To minimize the risk that we map the hearts of all individuals to a heart that is not representative or extreme in some relevant aspect, we aimed to select templates that reside close to the middle of the distribution of the subject age and heart region volumes (LVV, RVV, LAV, RAV, MV). We minimize the sum of weighted z-scores (after winsorizing/clipping each feature using percentiles 1 and 99 to avoid influence by extreme values), where age is assigned weight 1, and each of the volume measurements is assigned weights $\frac{1}{5}$ to place equal importance on age and volume. 

\begin{figure}[ht]
    \centering
    \subfloat[][Female: Age\label{fig:femaletemplateselection_a}]{\includegraphics[width=0.33\textwidth]{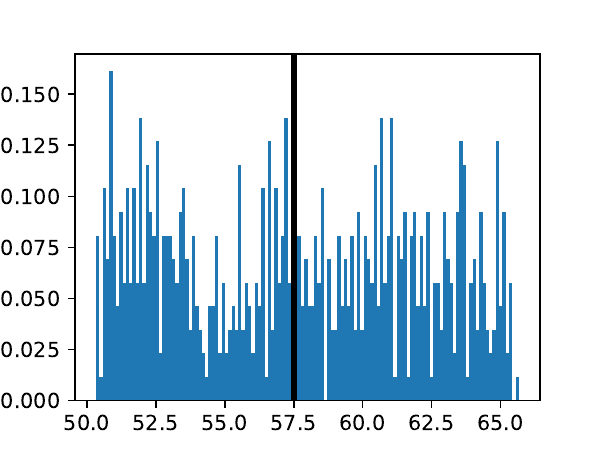}}
    \subfloat[][Female: LVV\label{fig:femaletemplateselection_b}]{\includegraphics[width=0.33\textwidth]{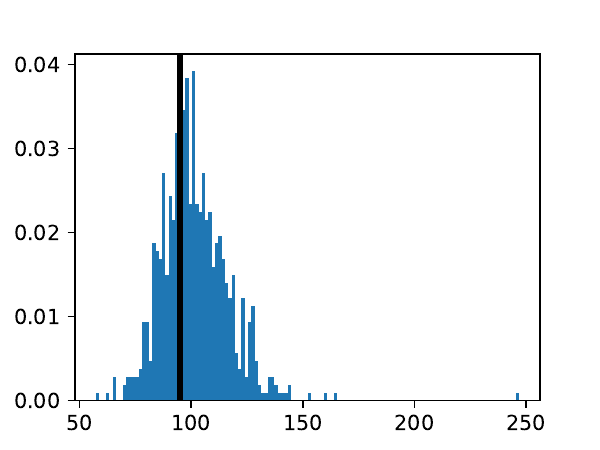}}
    \subfloat[][Female: RVV\label{fig:femaletemplateselection_c}]{\includegraphics[width=0.33\textwidth]{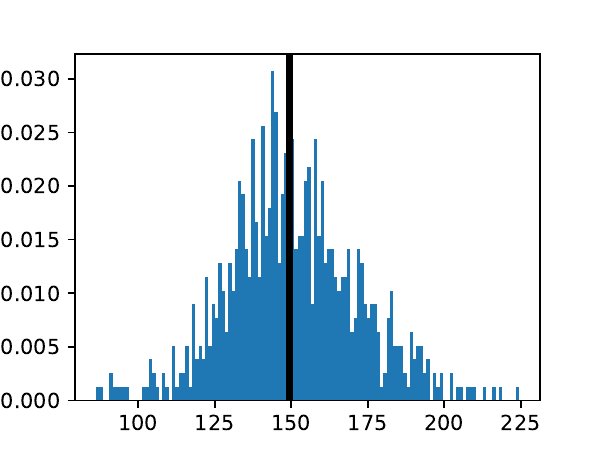}}\\
    \subfloat[][Female: LAV\label{fig:femaletemplateselection_d}]{\includegraphics[width=0.33\textwidth]{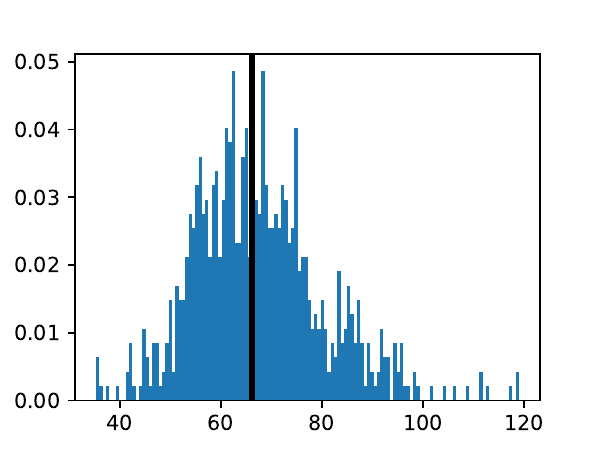}}
    \subfloat[][Female: RAV\label{fig:femaletemplateselection_e}]{\includegraphics[width=0.33\textwidth]{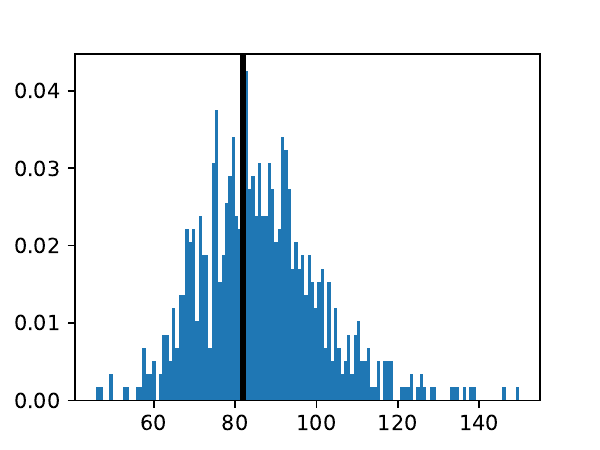}}
    \subfloat[][Female: MV\label{fig:femaletemplateselection_f}]{\includegraphics[width=0.33\textwidth]{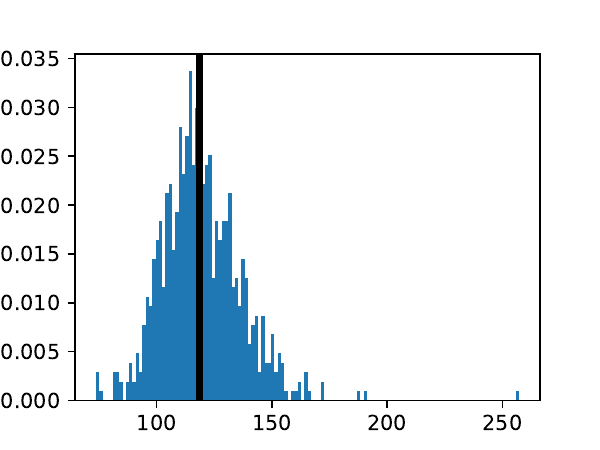}}
    \caption{Analysis of the template selection for the female cohort. The sub-figures (a-f) display the histograms of the features used to select the template with age measured in years and volumes measured in mL. The value of each feature in the selected template is shown as a black bar overlaying the histogram.}
    \label{fig:femaletemplateselection}
\end{figure}

\begin{figure}[ht]
    \centering
    \subfloat[][Male: Age\label{fig:maletemplateselection_a}]{\includegraphics[width=0.33\textwidth]{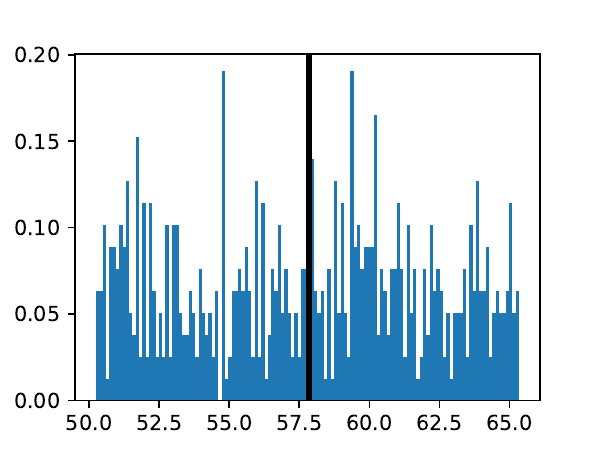}}
    \subfloat[][Male: LVV\label{fig:maletemplateselection_b}]{\includegraphics[width=0.33\textwidth]{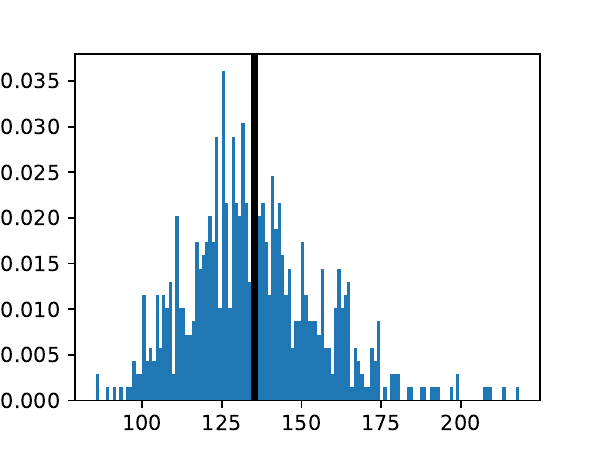}}
    \subfloat[][Male: RVV\label{fig:maletemplateselection_c}]{\includegraphics[width=0.33\textwidth]{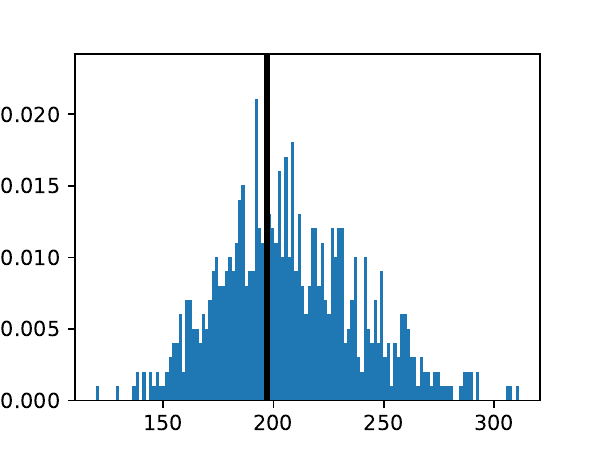}}\\
    \subfloat[][Male: LAV\label{fig:maletemplateselection_d}]{\includegraphics[width=0.33\textwidth]{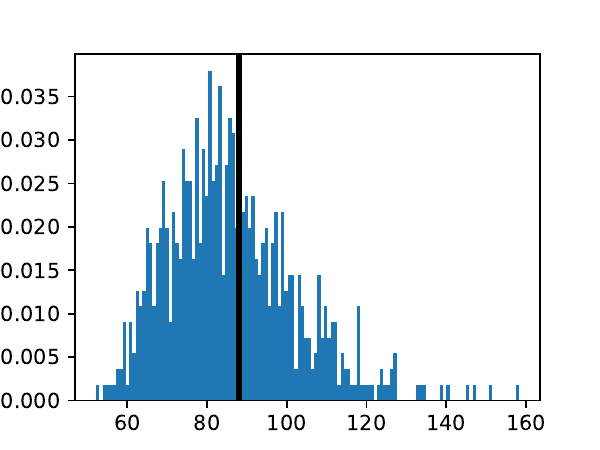}}
    \subfloat[][Male: RAV\label{fig:maletemplateselection_e}]{\includegraphics[width=0.33\textwidth]{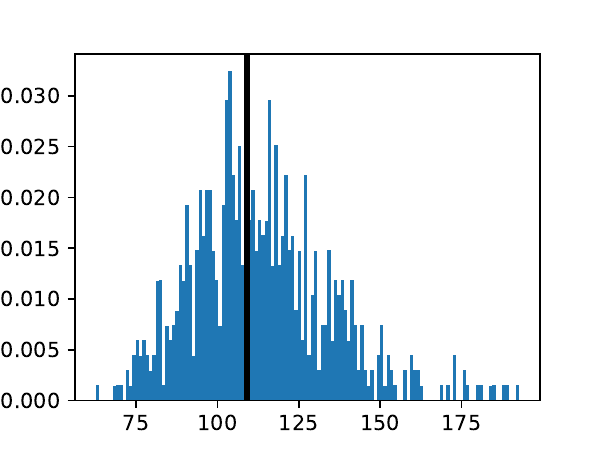}}
    \subfloat[][Male: MV\label{fig:maletemplateselection_f}]{\includegraphics[width=0.33\textwidth]{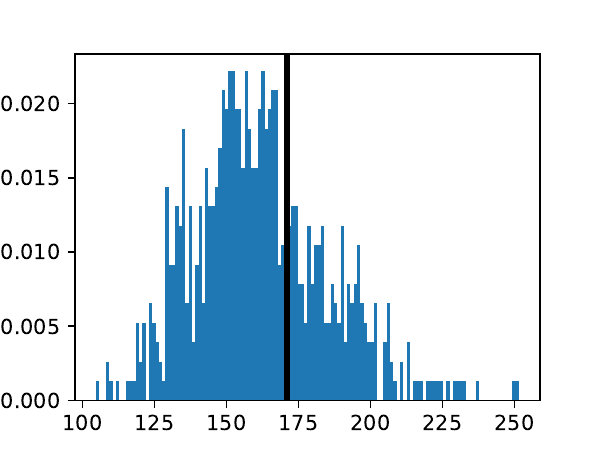}}
    \caption{Analysis of the template selection for the male cohort. The sub-figures (a-f) display the histograms of the features used to select the template with age measured in years and volumes measured in mL. The value of each feature in the selected template is shown as a black bar overlaying the histogram.}
    \label{fig:maletemplateselection}
\end{figure}

This process was repeated for males and females, yielding two templates, enabling sex-stratified analysis, which is of particular importance due to known substantial morphological differences between male and female hearts \citep{olivetti1995gender,st2022sex}. The statistics of the features used to select the template are shown for females in figures \ref{fig:femaletemplateselection_a}-\ref{fig:femaletemplateselection_f} and for males in figures \ref{fig:maletemplateselection_a}-\ref{fig:maletemplateselection_f}.

\subsubsection{Measures for Evaluation and Analysis of Deformation Fields}

Three main approaches were used to evaluate the image registration performance and analyze the deformation fields, in this work: (i) the Dice coefficient, (ii) the Jacobian Determinant (JD) of the deformation field, and (iii) the inverse consistency error.

The Dice coefficient \citep{dice1945measures} is a measure of overlap of two sets commonly used for evaluation of both image segmentation and image registration (through evaluation of how much known, segmented, structures overlap in the reference space). It is defined as
\begin{equation}
    \mathtt{DICE}(A, B) = 2\frac{\abs{A \cap B}}{\abs{A} + \abs{B}}.
\end{equation}

The JD of a deformation field measures the local expansion (where the JD>1) or contraction (where the JD < 1) \citep{yanovsky2007local}. Where the JD is negative, there are discontinuities in the deformation field, which tends to be undesirable in medical image analysis. The JD is used both for evaluation of the registration, through inspection of JD maps and JD standard deviation (SD) maps, and as a measure of local volume useful for supervoxel-wise studies of local volume changes in relation to a template.

Additionally, we explored the inverse consistency error \citep{christensen1999consistent} (ICE) of the registration process, which provides an estimate the smoothness and plausibility of the resulting transformations,
\begin{equation}
    \mathtt{ICE}(T^{R \rightarrow F}, T^{F \rightarrow R}) = \sum\limits_{x}||(T^{F \rightarrow R}(T^{R \rightarrow F}(x)) - x||_2,
\end{equation}
where $T^{R \rightarrow F}$ is the transformation from the reference space to a floating space, $T^{F \rightarrow R}$ is the transformation from the floating space to a reference space, and $x$ are the elements of the reference domain over which the measure is computed (the entire image space, a single element, or a subregion).

\subsubsection{Imiomics Analysis with Robust Statistics}

Imiomics (imaging-omics) is a methodology for evaluating the voxel-wise correlation between image values and non-imaging parameters of interest \citep{strand2017concept}, originally proposed for magnetic resonance imaging. It works in a two-step approach: (i) registration of all images into a common (template) space, and (ii) voxel-wise correlation/linear regression applied to the image values as the independent variable and a parameter of interest as the corresponding dependent variable. The attained correlation coefficients/regression slopes describe the relationship, and the $p$-values are used to identify voxels for which the analysis is statistically significant.

In this study, we applied Imiomics to both the density images (as measured in HU) and JD images which are obtained from the registration process, and which describe the local volumetric changes in relation to the template image, and thus enabling the study of morphology in relation to the selected target variable.

To exclude the display of statistical relationships in regions we de-emphasized during the image registration, we filtered out those regions (lungs, liver, stomach, esophagus) from the correlation maps, displaying them the same as insignificant regions.

\paragraph{Robustness and Computational Efficiency}

As we anticipated some (mostly local) registration failures, as well as outliers in intensity due to imaging artifacts or issues with inconsistent contrast enhancement, we employ supervoxel-wise outlier detection and rejection using the 1.5 Interquartile Range (IQR) method, discarding any values outside the interval
\begin{equation}
    \left[q_1 - 1.5 (q_3-q_1), q_3 + 1.5 (q_3-q_1)\right],
\end{equation}
where $q_1$ and $q_3$ denotes the first and third quartiles of the value.

To reduce the number of voxels that we need to store and process (to speed up the analysis and reduce the auxiliary memory required) and to reduce the number of statistical tests that we perform (to make the analysis more reliable) we propose to use supervoxels (applied to the template images). In this work, we used \emph{Simple Linear Iterative Clustering} (SLIC) \citep{achanta2012slic}, which has shown to be fast, robust, and follows natural object boundaries well. We then performed the supervoxel-wise regression on the mean values of each supervoxel, rather than on every voxel of the original image. Supervoxels have been successfully used in other works, e.g. association studies in neuroscience \citep{batmanghelich2013joint}.

\begin{figure}[ht]
    \centering
    \begin{turn}{90}
        \;\;\;\;\;\;\;\;\;\;\;\;\;Female       
    \end{turn}\hfill
    \subfloat[][SLIC: Slice 1]{\includegraphics[width=0.3\textwidth]{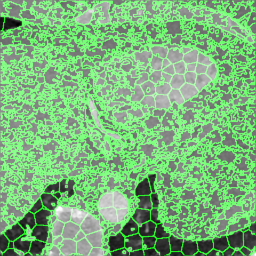}}\hfill
    \subfloat[][SLIC: Slice 2]{\includegraphics[width=0.3\textwidth]{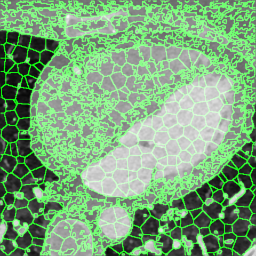}}\hfill
    \subfloat[][SLIC: Slice 3]{\includegraphics[width=0.3\textwidth]{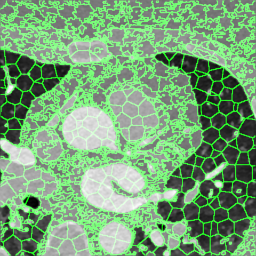}}
    \\
    \begin{turn}{90}
        \;\;\;\;\;\;\;\;\;\;\;\;\;\;Male       
    \end{turn}\hfill
    \subfloat[][SLIC: Slice 1]{\includegraphics[width=0.3\textwidth]{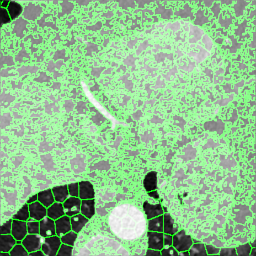}}\hfill
    \subfloat[][SLIC: Slice 2]{\includegraphics[width=0.3\textwidth]{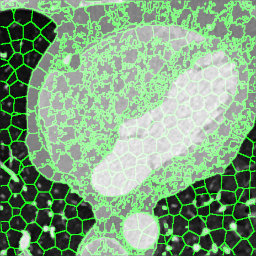}}\hfill
    \subfloat[][SLIC: Slice 3]{\includegraphics[width=0.3\textwidth]{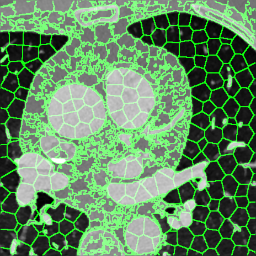}}    
    \caption{Illustration of the SLIC supervoxels (volumetric connected segments), as used in this work for a more robust and efficient Imiomics analysis, here shown on top of selected slices of both the female and male reference images. The large homogeneous regions of the ventricles, the vessels, and the aorta have more regularly shaped supervoxels given that there is less variation in density to influence the shape, while the other parts have much more high-frequency variations in the boundaries.}
    \label{fig:slicexample}
\end{figure}

We extracted supervoxels, using the SimpleITK implementation of SLIC \citep{lowekamp2018scalable}, from the template images by configuring SLIC to place seeds at every 25:th voxel along each dimension, with spatial proximity weight set to 0.2, and forcing connectedness of the clusters. An example of the SLIC supervoxels computed for the template images is shown in Fig.~\ref{fig:slicexample}. For each cluster, we computed the statistics of the density and JD images within each cluster. To reduce the effects of registration errors and effects of supervoxels containing tissues of different densities, we applied a filter where voxels with a density (HU) outside an interval given by the 1.5 IQR outlier detection method were excluded from the mean value calculation for both the density images and JD images. 

\paragraph{Proof-of-Concept Analysis - Measuring the Correlation to Volumetric Parameters with Known Spatial Localization}

To establish confidence in the complete pipeline from pre-processing, image registration, supervoxel extraction, and statistical analysis, we performed a proof-of-concept study where supervoxel-wise analysis is done between the sex-stratified stacks of JD images and volumetric measurements (LVV and LAV) derived from the images using image segmentation with TotalSegmentator.

\paragraph{Association of Age to Volume and Density}

We studied the relationship between the independent variables of density and volume (JD) against the dependent variable of age, with supervoxel-wise Pearson correlation.

\section{Results}

\subsection{Evaluation of the Image Registration}

\begin{figure}[ht]
    \centering
    \begin{turn}{90} 
        \;\;\;Female       
    \end{turn}\hfill
    \subfloat[][Reference]{\includegraphics[width=0.15\textwidth]{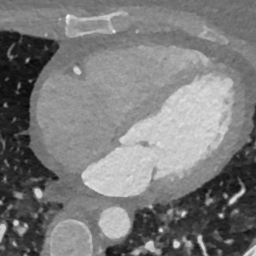}}\hfill
    \subfloat[][HU Mean]{\includegraphics[width=0.15\textwidth]{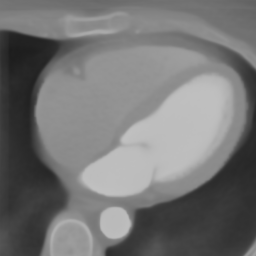}}\hfill
    \subfloat[][HU SD]{\includegraphics[width=0.15\textwidth]{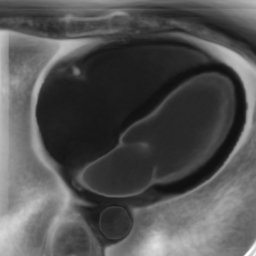}}\hfill
    \subfloat[][JD Mean]{\includegraphics[width=0.15\textwidth]{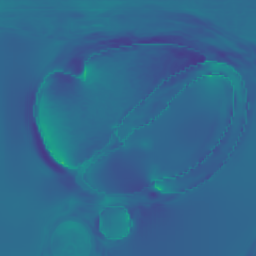}}\hfill
    \subfloat[][JD SD]{\includegraphics[width=0.15\textwidth]{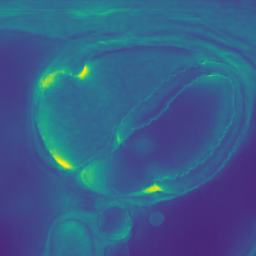}}\hfill
    \subfloat[][ICE Mean]{\includegraphics[width=0.15\textwidth]{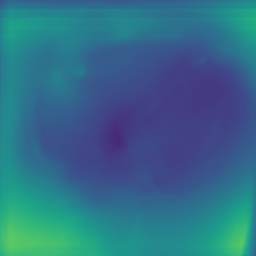}}    
    \\
    \begin{turn}{90} 
        \;\;\;\;\;Male       
    \end{turn}\hfill
    \subfloat[][Reference]{\includegraphics[width=0.15\textwidth]{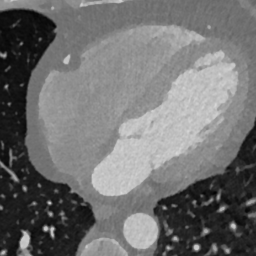}}\hfill    \subfloat[][HU Mean]{\includegraphics[width=0.15\textwidth]{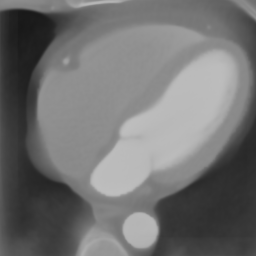}}\hfill
    \subfloat[][HU SD]{\includegraphics[width=0.15\textwidth]{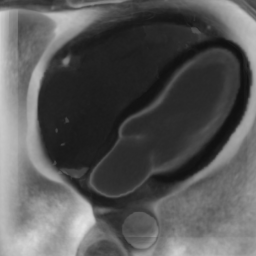}}\hfill
    \subfloat[][JD Mean]{\includegraphics[width=0.15\textwidth]{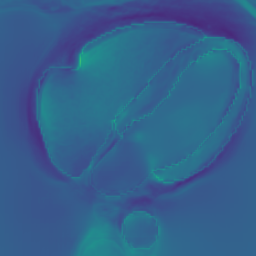}}\hfill
    \subfloat[][JD SD]{\includegraphics[width=0.15\textwidth]{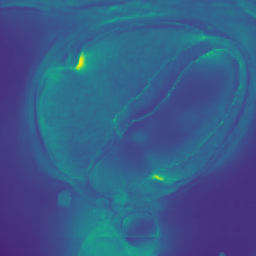}}\hfill
    \subfloat[][ICE Mean]{\includegraphics[width=0.15\textwidth]{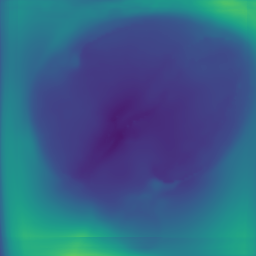}}\\
    \begin{turn}{90} 
        CB       
    \end{turn}\hfill
    \includegraphics[width=0.15\textwidth]{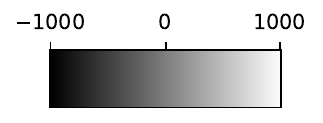}\hfill
    \includegraphics[width=0.15\textwidth]{figs/colorbars/ct_cbar.pdf}\hfill
    \includegraphics[width=0.15\textwidth]{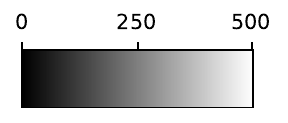}\hfill
    \includegraphics[width=0.15\textwidth]{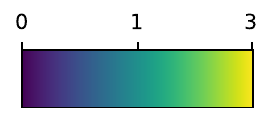}\hfill
    \includegraphics[width=0.15\textwidth]{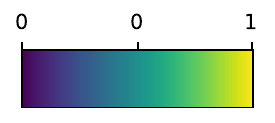}\hfill
    \includegraphics[width=0.15\textwidth]{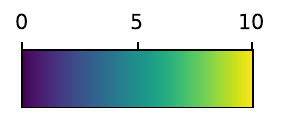}
    \caption{Visual examples of the registration performance showing example axial slices of: (a, g) the reference images, (b,h)  the voxel-wise mean volume corresponding to the female cohort ($n=722$) and male cohort ($n=666$), (c, i) the voxel-wise SD images, (d, j) the voxel-wise mean JD, (e, k) the voxel-wise SD of the JD, (f, l) the voxel-wise ICE error. The main structures of the heart (chambers, myocardium, aorta) appear well formed and aligned, while there are some regions consisting of fat and coronal vessel structures that were not consistently well aligned, and thus appearing as fuzzy in the mean image, with higher standard deviation, and differences in JD images, as well as in the ICE images. JD and ICE are presented using a colormap, where lower, cooler colors indicate smaller errors/better performance. We observe that the overall ICE is low within the heart, and higher outside where the lungs were removed from the CT images, as seen in Fig.~\ref{fig:registrationillustration}, and in other surrounding tissue.}
    \label{fig:heartregistrationexample}
\end{figure}

In Fig.~\ref{fig:heartregistrationexample} we display the mean and standard deviation images after registration, as well as the mean and standard deviation Jacobian determinant images, and the voxel-wise ICE images, for both male and female to enable a visual evaluation of the performance of the registration. See \ref{sec:appendixregistrations} for additional examples of axial slices, with corresponding visualizations.

\begin{figure}[ht]
    \centering
    \subfloat[][Registration performance (Dice score) on the female cohort.]{\includegraphics[width=0.49\textwidth]{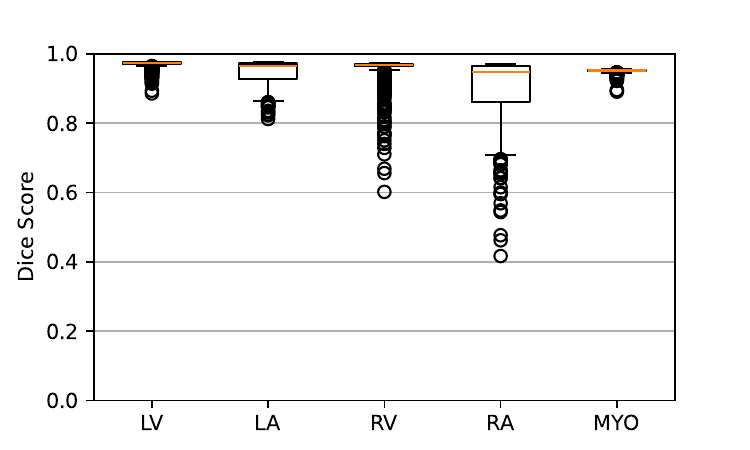}} \hfill
    \subfloat[][Registration performance (Dice score) on the male cohort.]{\includegraphics[width=0.49\textwidth]{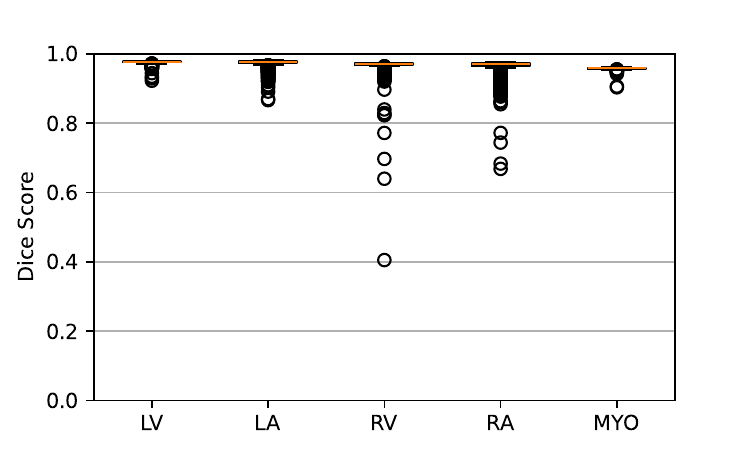}}\\ 
    \subfloat[][Registration performance (ICE in \texttt{mm}) on the female cohort.]{\includegraphics[width=0.49\textwidth]{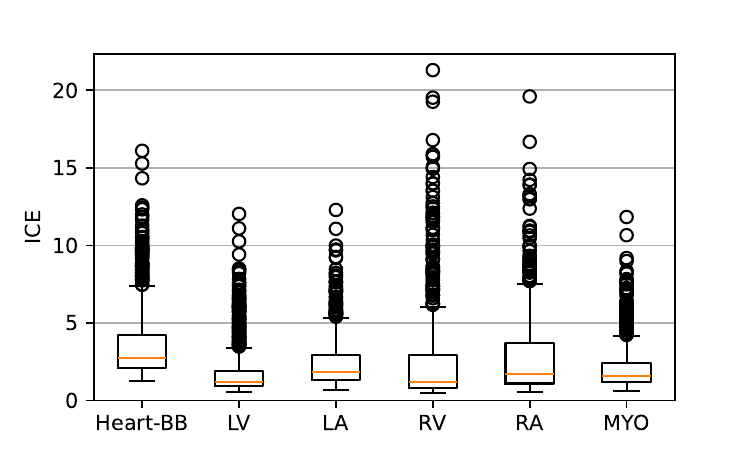}} \hfill
    \subfloat[][Registration performance (ICE in \texttt{mm}) on the male cohort.]{\includegraphics[width=0.49\textwidth]{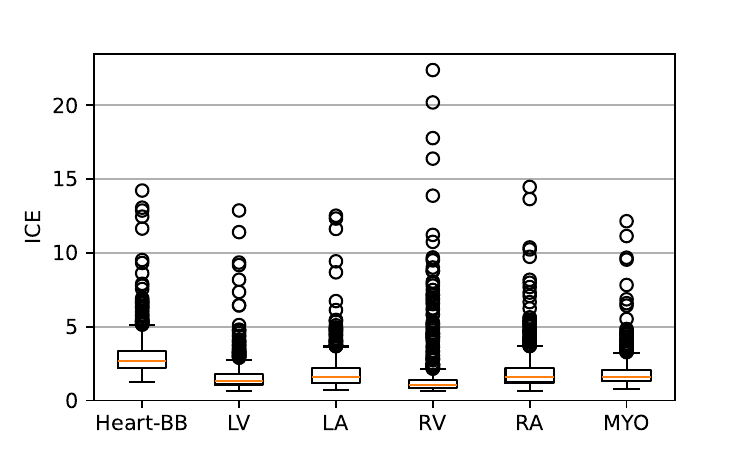}}\\    
    \caption{Performance of image registration in terms of the Dice coefficient of five major parts of the heart (LV, RV, LA, RA, MYO), for (a) female subjects and (b) male subjects, and ICE evaluated within regions of interest (Heart-BB, LV, LA, RV, RA, MYO), for (c) female subjects, and (d) male subjects. The performance is excellent in terms of the Dice score for most of the individuals on the left side (chambers and the myocardium). We observe very few failure cases where these anatomical regions do not overlap with at least a majority of their volume. The performance is good overall on the right side (RV and RA) even though worse than on the left side, with more outliers, as well as a wide IQR for RA in the female cohort. We observed an ICE of around a few \texttt{mm} in all regions, for most subjects. A number of outliers for each region exhibit a much higher ICE, indicating more substantial inconsistencies.}
    \label{fig:regperfdice}
\end{figure}

Figure~\ref{fig:regperfdice} shows the performance of the image registration in terms of overlap of the segmentation masks (LV, LA, RV, RA, MYO) measured using the Dice coefficient, presented as box plots for males and females separately. We also present numerical aggregates corresponding to these results in Tab.~\ref{tab:regresults}. We observed high performance for most of the subjects (and regions), for all five segments, with higher performance on the left side (LV, LA, MYO) than on the right (RV, RA). In particular, the female cohort exhibited a number of outliers on the right side, whereas the male cohort had mostly high performance with a low number of outliers.

\begin{table}[]
    \centering
    \caption{Quantitative evaluation of the registration method in terms of the Dice score and ICE within the five main sub-regions of the heart that were evaluated.}
    \label{tab:regresults}
    \resizebox{\columnwidth}{!}{%
    \begin{tabular}{c|c|ccccc}
        Sex & Measure & LV & LA & RV & RA & MYO \\ \hline
        \multirow{2}{*}{Female} & Dice & $0.970 \pm 0.011$ & $0.947 \pm 0.033$ & $0.956 \pm 0.041$ & $0.904 \pm 0.086$ & $0.951 \pm 0.006$ \\
         & ICE (mm) & $1.90 \pm 1.71$ & $2.46 \pm 1.73$ & $2.64 \pm  3.23$ & $2.88 \pm 2.74$ & $2.15 \pm 1.60$ \\\hline
        \multirow{2}{*}{Male} & Dice & $0.977 \pm 0.005$ & $0.973 \pm 0.013$ & $0.967 \pm 0.031$ & $0.961 \pm 0.028$ & $0.958 \pm 0.003$ \\
        & ICE (mm) & $1.61 \pm 1.09$ & $1.91 \pm 1.21$ & $1.64 \pm 2.09$ & $1.95 \pm 1.36$ & $1.87 \pm 1.07$
    \end{tabular}}
\end{table}

One limitation of the evaluation of overlap (with the Dice score), is that we have used these masks to perform the registration, so they are not independent evaluation targets. On the other hand, by observing their overlap, and at the same time ascertaining that the deformations are plausible, in terms of the JD, overall mean density, and inverse consistency, this evaluation provides a check if the main structures were overall well-registered, and plausible.

\subsection{Pairwise Correlations Between Explicit Volumetric and Density Measurements Based on Image Segmentation}

\begin{figure}[ht]
    \centering
    \includegraphics[width=1\textwidth]{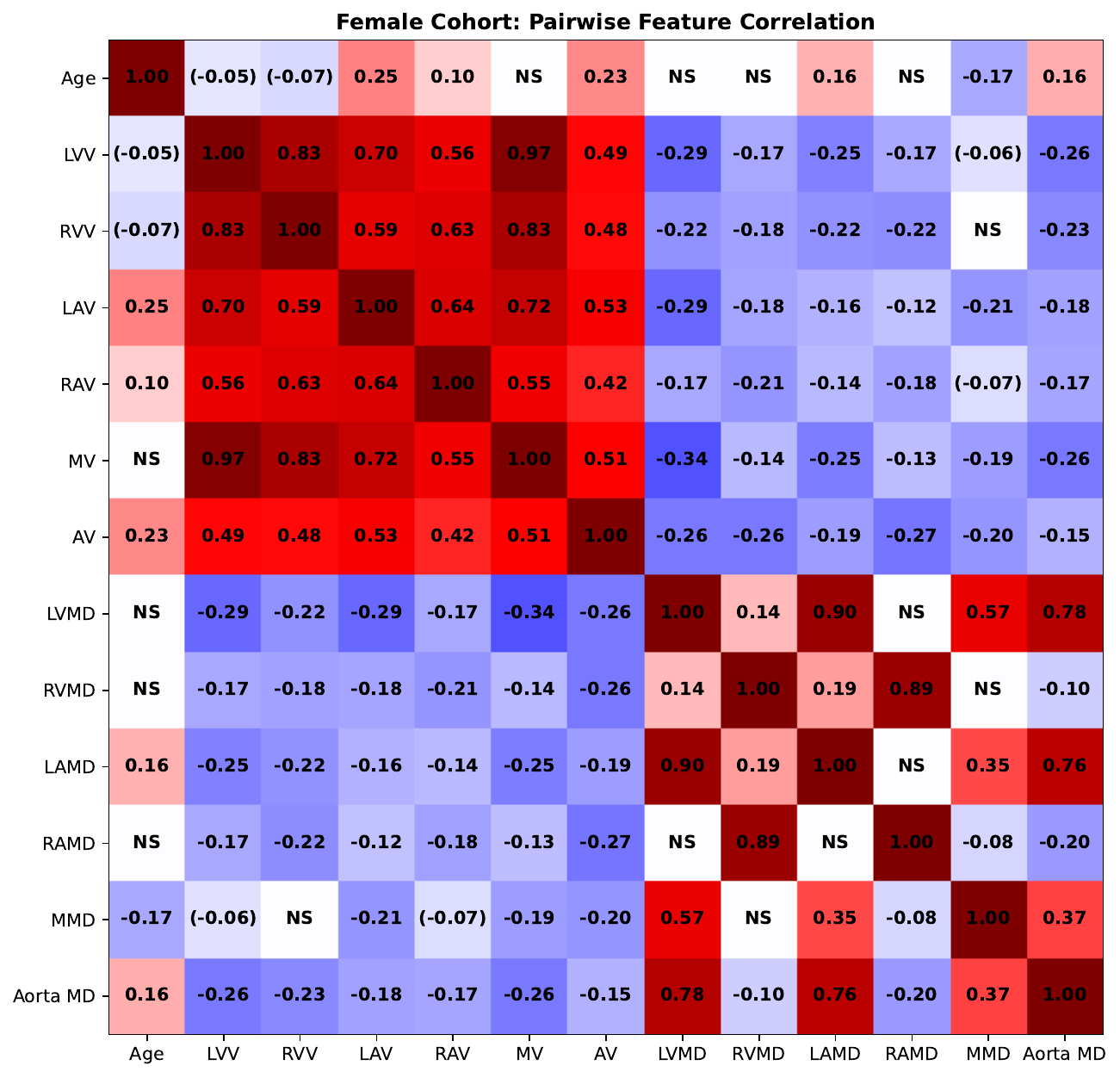}\\
    \hfill\includegraphics[width=0.3\textwidth]{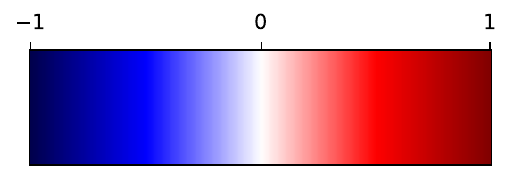}\\    \caption{Pairwise Pearson correlation coefficients between age and explicit volumetric measurements for the female cohort. NS denotes that the p-value was $>0.2$ and therefore omitted. A value enclosed in parentheses means that it is close to significance (0.05-0.2). All other values are statistically significant with $p<0.05$. (LVV: Left ventricle volume; RVV: Right ventricle volume; LAV: Left atrium volume; RAV: Right atrium volume; MV: Myocardium volume; AV: Aorta volume; LVMD: Left ventricle mean density; LAMD: Left atrium mean density; RVMD: Right ventricle mean density; RAMD: Right atrium mean density; MMD: Myocardium mean density; Aorta MD: Aorta mean density.)}
    \label{fig:pairwisecorrelations}
\end{figure}

\begin{figure}[ht]
    \centering
    \includegraphics[width=1\textwidth]{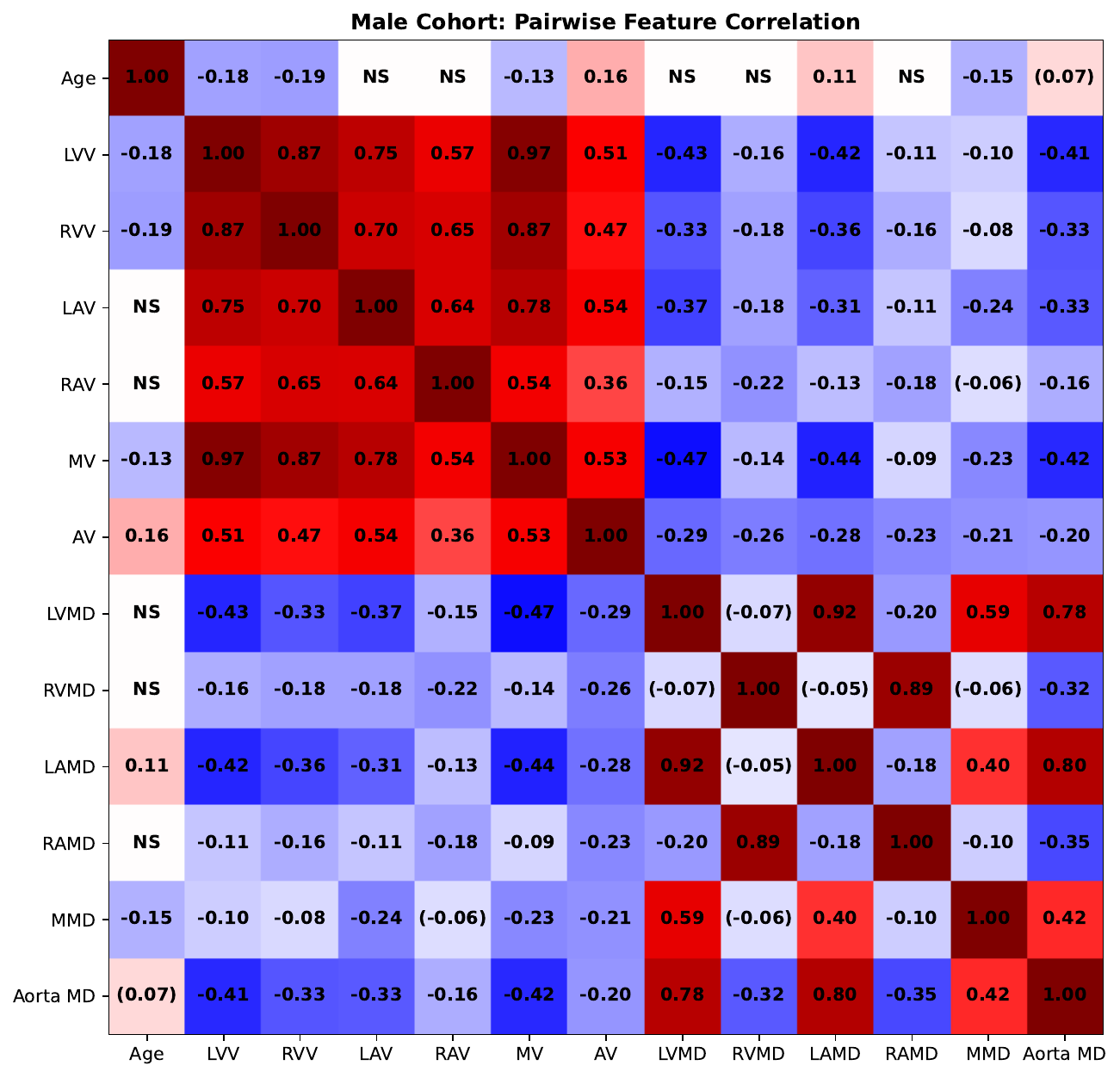}\\
    \hfill\includegraphics[width=0.3\textwidth]{figs/colorbars/correlation_cbar.pdf}\\    \caption{Pairwise Pearson correlation coefficients between age and explicit volumetric measurements for the male cohort. NS denotes that the p-value was $>0.2$ and therefore omitted. A value enclosed in parentheses means that it is close to significance (0.05-0.2). All other values are statistically significant with $p<0.05$. (LVV: Left ventricle volume; RVV: Right ventricle volume; LAV: Left atrium volume; RAV: Right atrium volume; MV: Myocardium volume; AV: Aorta volume; LVMD: Left ventricle mean density; LAMD: Left atrium mean density; RVMD: Right ventricle mean density; RAMD: Right atrium mean density; MMD: Myocardium mean density; Aorta MD: Aorta mean density.)}
    \label{fig:pairwisecorrelations2}
\end{figure}

We measured the pairwise correlations, in terms of the Pearson correlation coefficient, between age and volumetric (as well as mean density) features, using volumes measured using the segmented images from TS. The correlations for females are presented in Fig.~\ref{fig:pairwisecorrelations} and for males in Fig.~\ref{fig:pairwisecorrelations2}. We observe that the LVV correlates negatively with age in both males, which is in agreement with \citep{bai2020population} and \citep{wasserthal2023totalsegmentator}, and females (although not statistically significant for the common threshold $p<0.05$). In females, the LAV correlates positively with age, unlike for men (where we do not observe statistical significance). This observed association is in agreement with \citep{wasserthal2023totalsegmentator}, while not in agreement with the results observed in \citep{bai2020population}. The LAV measured in  \citep{bai2020population} is, however, estimated differently than in this study and in \citep{wasserthal2023totalsegmentator}.

\clearpage

\subsection{Supervoxel-wise Assoication Analysis}

First, we show the results of the proof-of-concept study, associating explicitly measured volumes with the JD, to evaluate the performance of the registration method and thereby the ability of the method to pick up associations we have prior knowledge about, and then we proceed to show the results of the associations with age.

\subsubsection{Proof-of-Concept Analysis: Measuring the Correlation to Volumetric Parameters with Known Spatial Localization}

\begin{figure}[ht]
    \centering    
    \textbf{Supervoxel-wise associations between local volume (JD) and explicit volume measurements as a proof-of-concept}\\
    \begin{turn}{90} 
        \;\;\;Female       
    \end{turn}
    \subfloat[][Supervoxel-wise Pearson correlation between the local volume and LAV. Maximum correlation: 0.88.]{\includegraphics[width=0.47\textwidth]{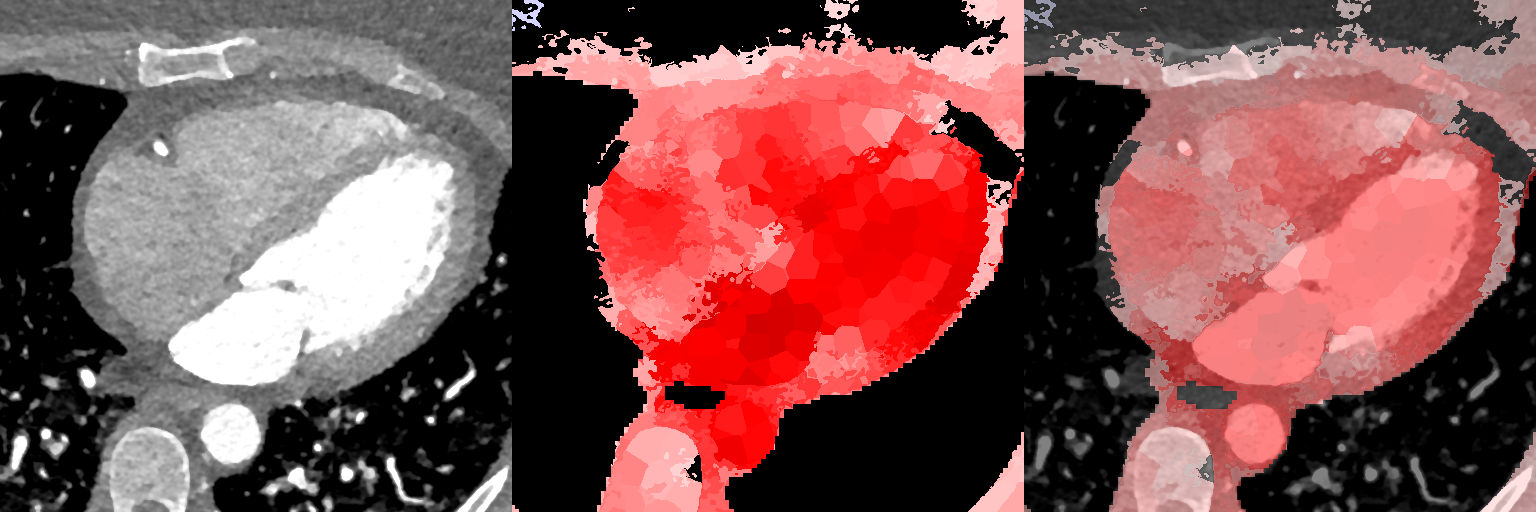}} \hfill
    \subfloat[][Supervoxel-wise Pearson correlation between the local volume and LVV. Maximum correlation: 0.87.]{\includegraphics[width=0.47\textwidth]{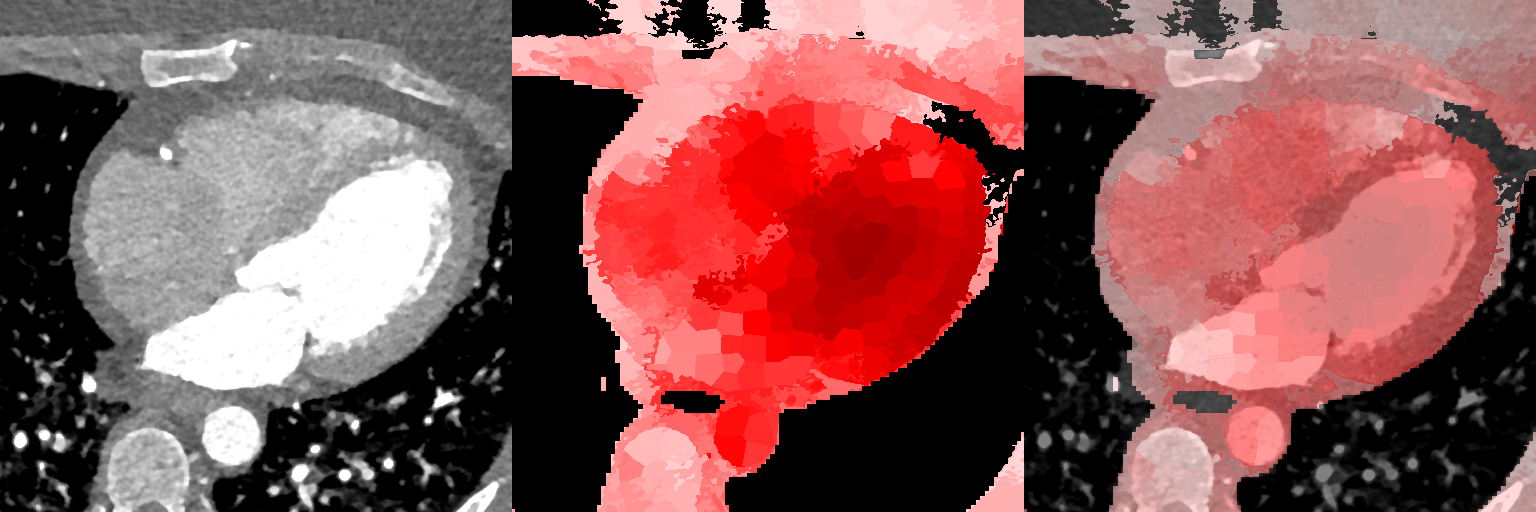}}    
    \\
    \begin{turn}{90} 
        \;\;\;\;\;Male       
    \end{turn}
    \subfloat[][Supervoxel-wise Pearson correlation between the local volume and LAV. Maximum correlation: 0.91.]{\includegraphics[width=0.47\textwidth]{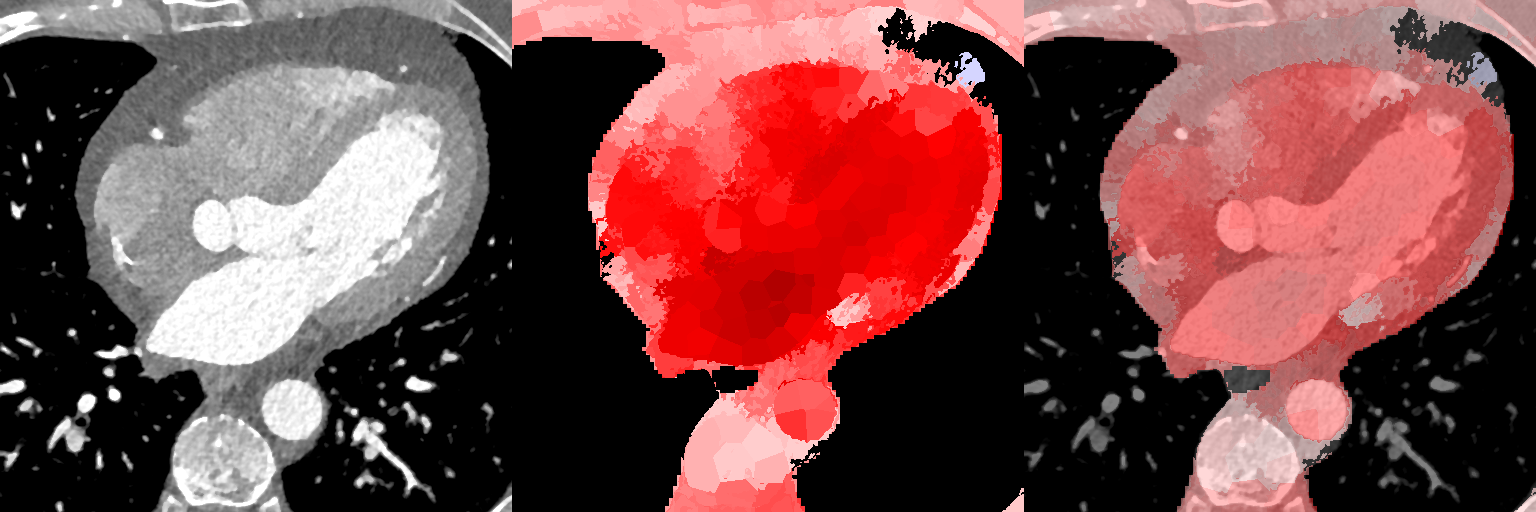}} \hfill
    \subfloat[][Supervoxel-wise Pearson correlation between the local volume and LVV. Maximum correlation: 0.88.]{\includegraphics[width=0.47\textwidth]{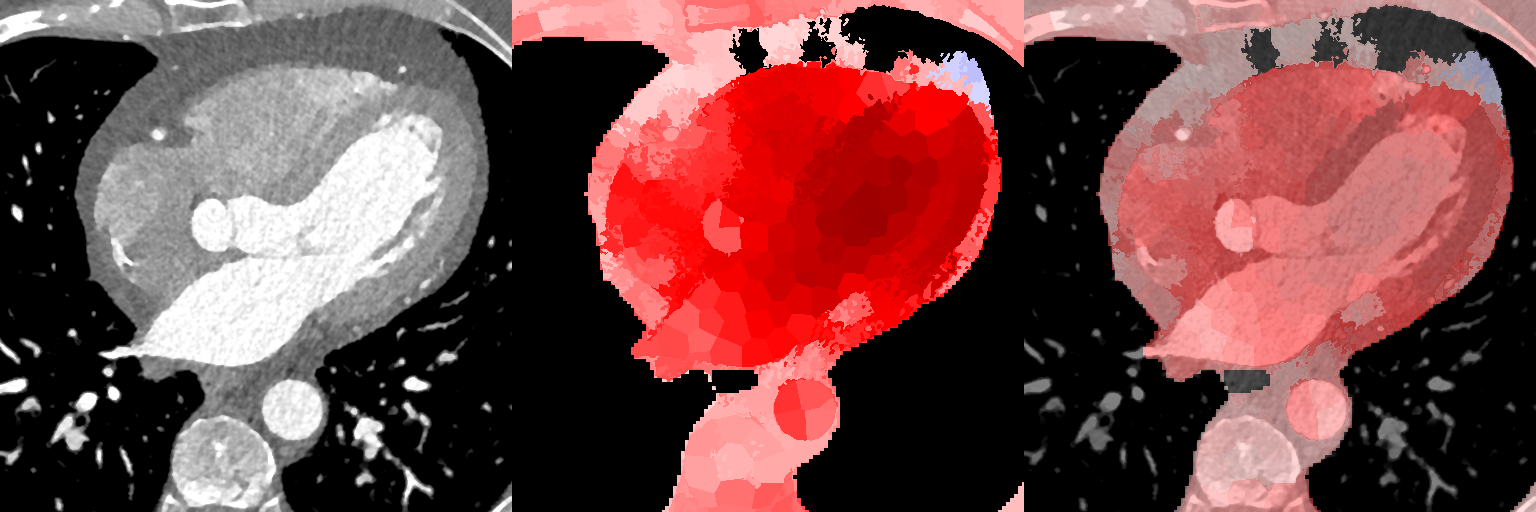}}\\
    \hfill\includegraphics[width=0.3\textwidth]{figs/colorbars/correlation_cbar.pdf}\\    
    \caption{Supervoxel-wise results of Pearson correlation between JD and LAV/LVV, here illustrated on a single slice for each variable and sex. Each image is displayed as a reference image slice, correlation maps (where non-significance is shown as black), and the correlation map overlaid on the reference image. High maximum correlations between the JD and the corresponding explicit measurements are detected within both the LA and LV regions.}
    \label{fig:pocimiomics}
\end{figure}

Figure \ref{fig:pocimiomics} displays slices from the supervoxel-wise maps encoding how the local volume deformations correlate with explicit volume measures. The ideal outcome is a correlation coefficient of one within the entire region being measured. We observe a high (while less than ideal) maximal correlation of between 0.85-0.91 for the LVV and LAV variables, and observe a variation of correlation strength over the extent of the regions being measured, indicating that the registration and the supervoxel-wise association analysis works well, but not perfectly and that the registration has deformed these regions non-uniformly to a limited extent.

\subsubsection{Supervoxel-wise Association with Age}

\begin{figure}[ht]
    \centering
    \textbf{Supervoxel-wise associations between local volume (JD) and age} \\
    \begin{turn}{90} 
        \;\;\;Female       
    \end{turn}
    \subfloat[][Reference, Pearson correlation, Pearson Correlation overlaid on reference, for two selected axial slices of the female cohort.]{\includegraphics[width=0.47\textwidth]{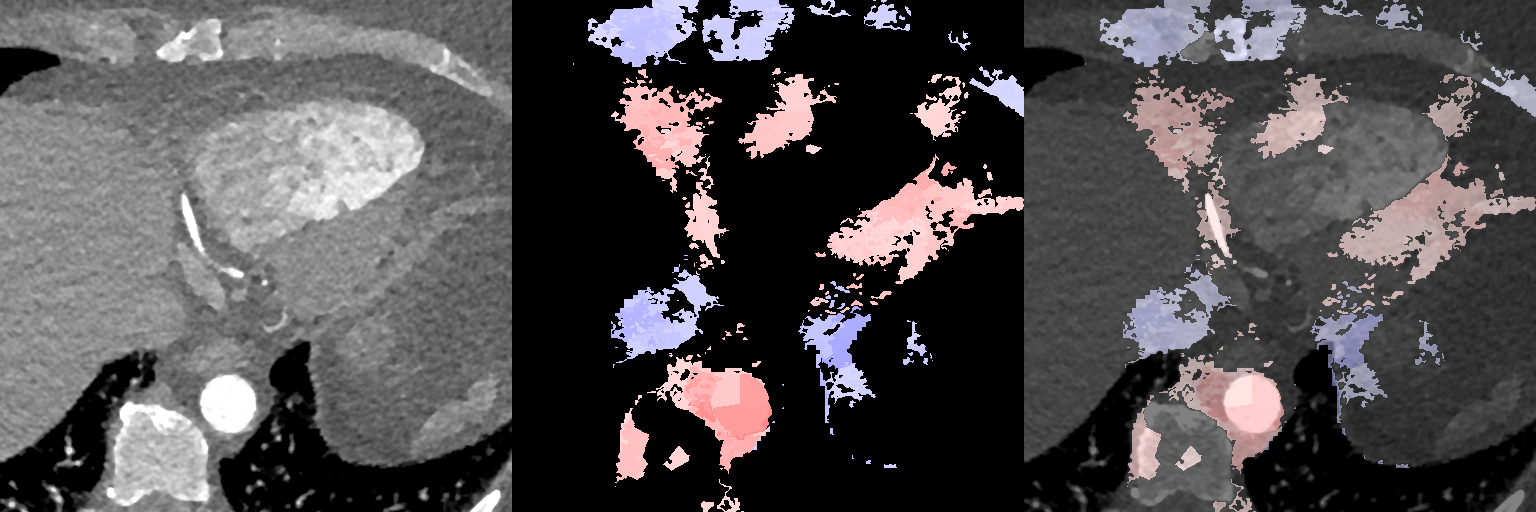} \hfill
    \includegraphics[width=0.47\textwidth]{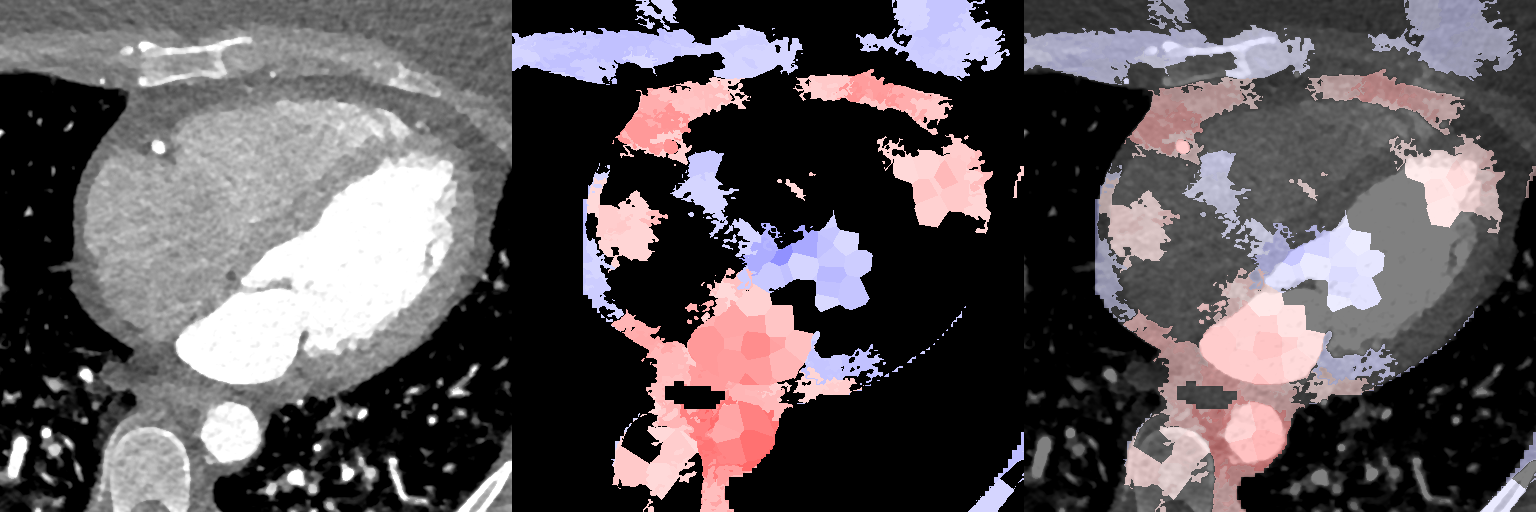}}\\
    \begin{turn}{90} 
        \;\;\;\;\;Male       
    \end{turn}
    \subfloat[][Reference, Pearson correlation, Pearson Correlation overlaid on reference, for two selected axial slices of the male cohort.]{\includegraphics[width=0.47\textwidth]{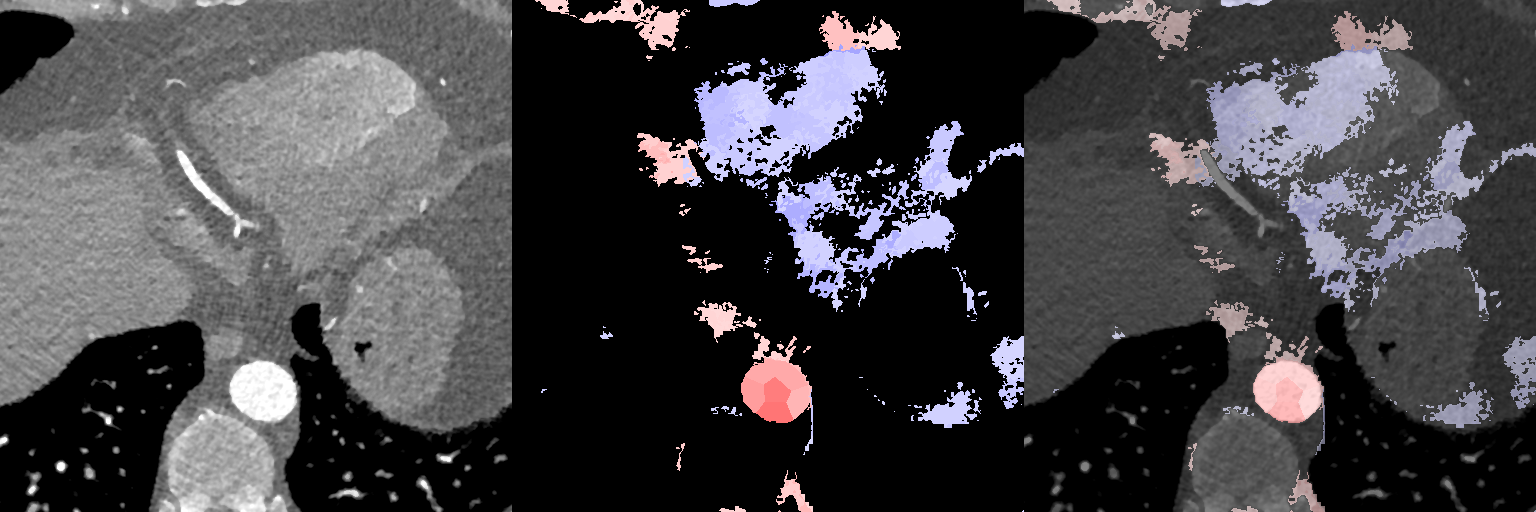} \hfill
    \includegraphics[width=0.47\textwidth]{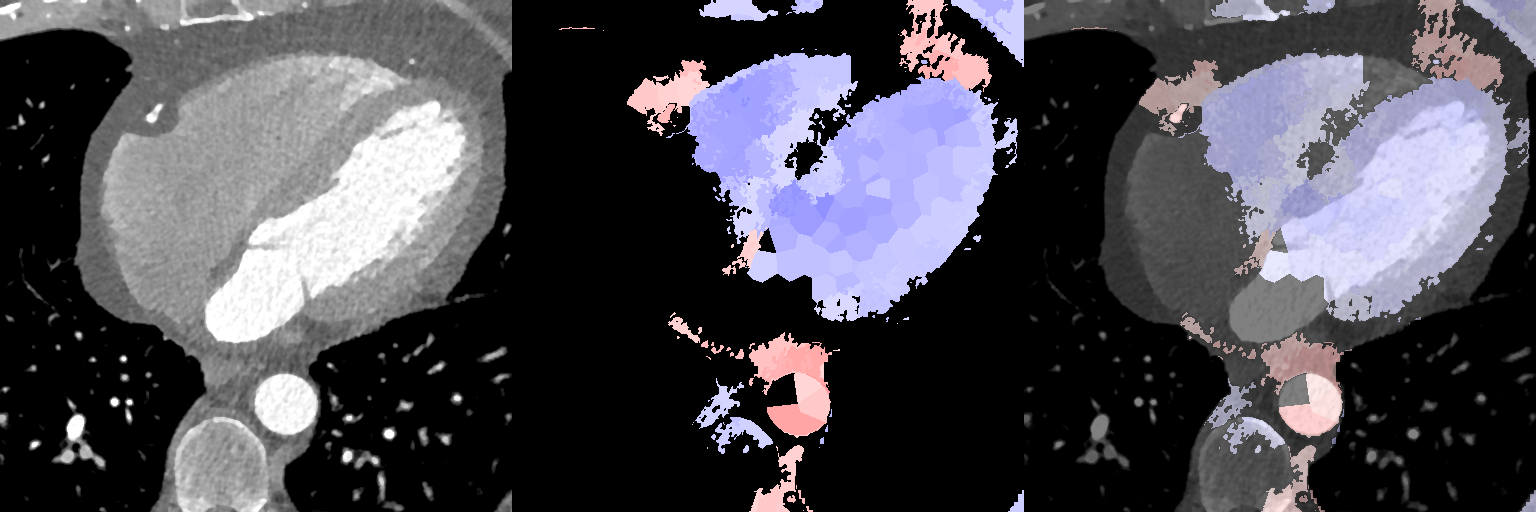}}\\
    \hfill\includegraphics[width=0.3\textwidth]{figs/colorbars/correlation_cbar.pdf}\\    
    \caption{Supervoxel-wise analysis of the local voxel volume, as measured by the JD, and the chronological age of the subject. Each image is displayed as a reference image slice, correlation maps (where non-significance is shown as black), and the correlation map overlaid on the reference image. We observe that the LAV has a positive correlation with age in females and that there is a positive correlation between the volume of the apex in females, and not in males. The LVV and RVV have a negative correlation with age (in small sub-regions in females), and the aorta volume exhibits a positive correlation with age.}
    \label{fig:imiomics_jd_age_results}
\end{figure}

In Fig.~\ref{fig:imiomics_jd_age_results}, we show association maps between age and local volume. We observe that the local volume correlates negatively with age in the LV, in agreement with the pairwise correlation study. We observe that females have a significant positive correlation between the JD in the LA and the chronological age of the subjects, while the male cohort is not statistically significant in the LA (having an overall weak positive correlation with a maximum $p$-value 0.08 in a single supervoxel, and above 0.2 in most supervoxels in the LA). The volume of the aorta (ascendens and descendens) are both positively correlated with age.

\begin{figure}[ht]
    \centering
    \textbf{Supervoxel-wise associations between density (HU) and age} \\
    \begin{turn}{90} 
        \;\;\;Female       
    \end{turn}
    \subfloat[][Reference, Pearson correlation, Pearson Correlation overlaid on reference, respectively, for two selected axial slices of the female cohort.]{\includegraphics[width=0.47\textwidth]{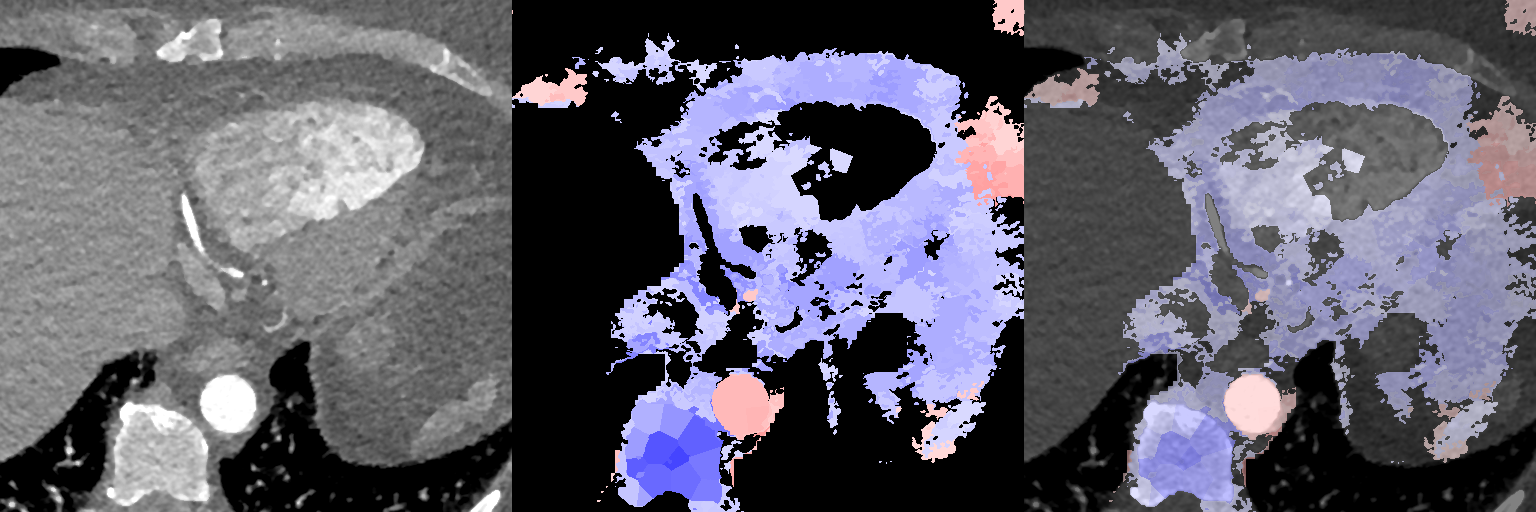}\;
    \includegraphics[width=0.47\textwidth]{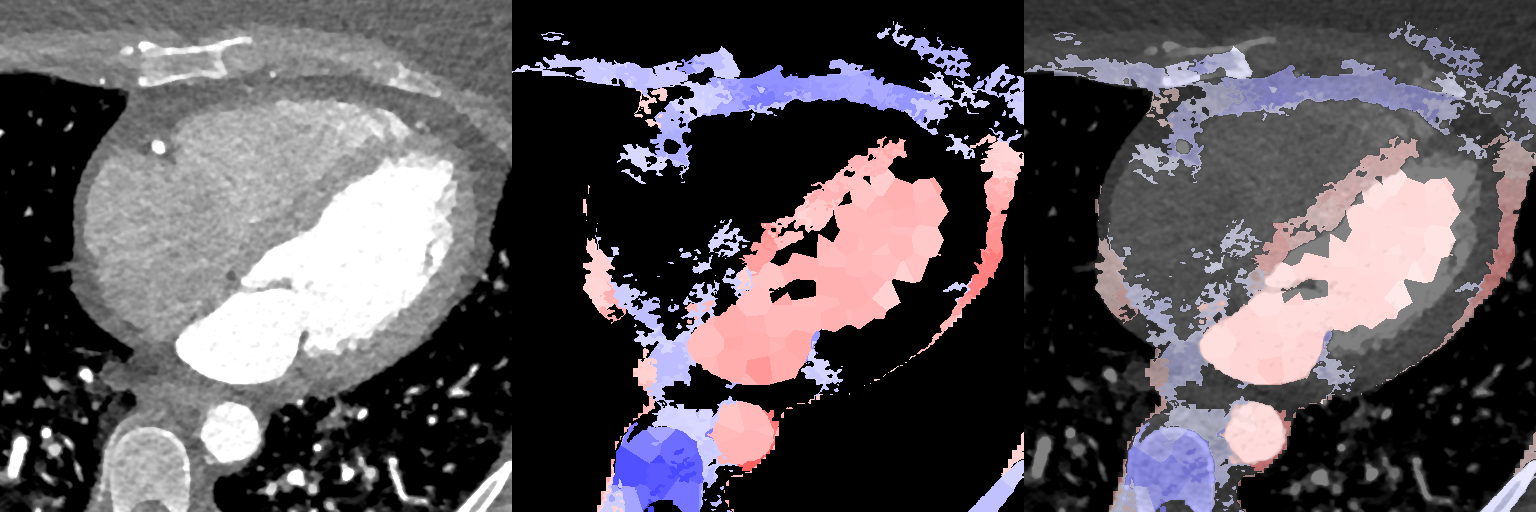}}\\
    \begin{turn}{90} 
        \;\;\;\;\;Male       
    \end{turn}
    \subfloat[][Reference, Pearson correlation, Pearson Correlation overlaid on reference, for two selected axial slices of the male cohort.]{\includegraphics[width=0.47\textwidth]{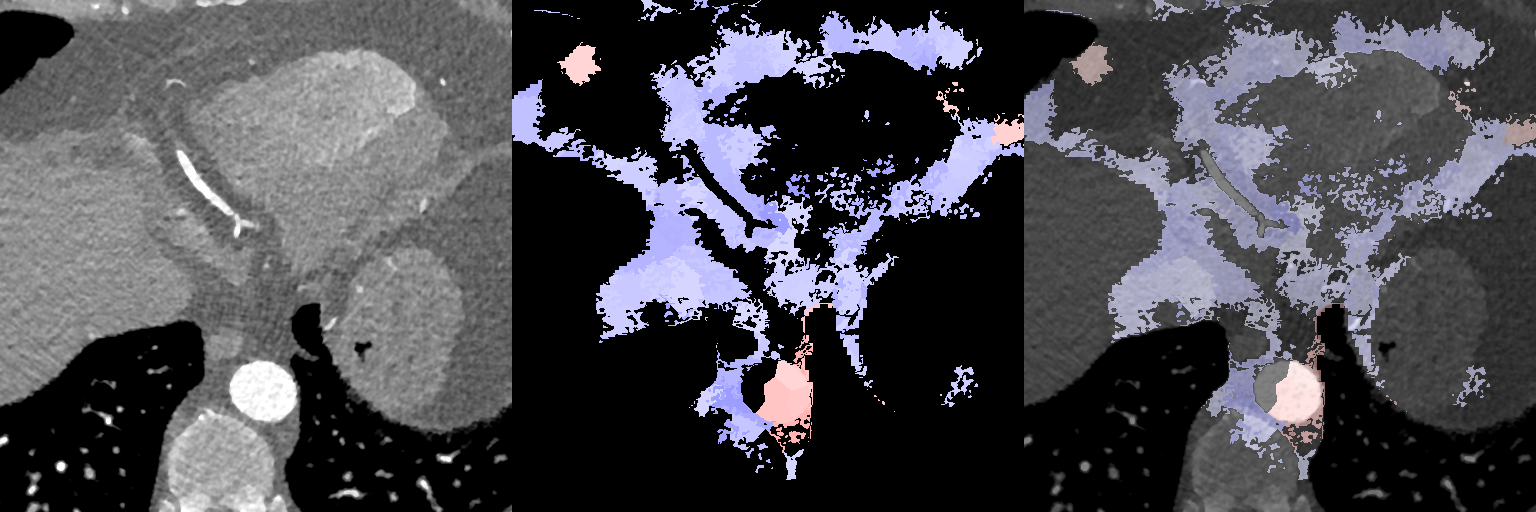}\;
    \includegraphics[width=0.47\textwidth]{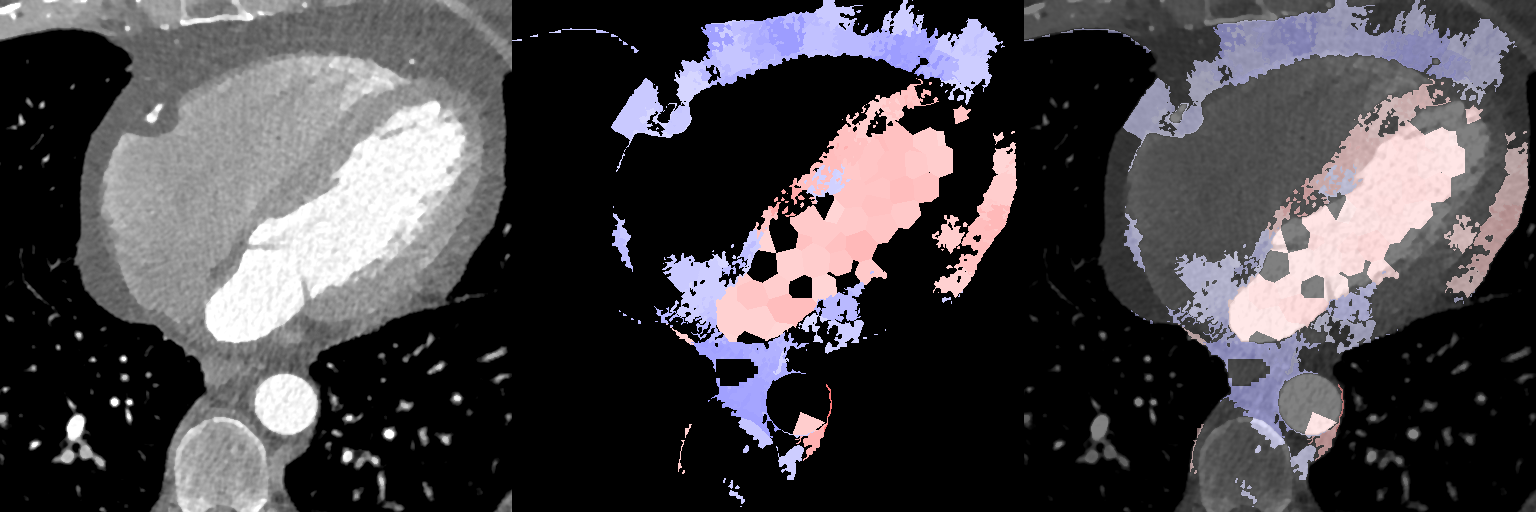}}\\
    \hfill\includegraphics[width=0.3\textwidth]{figs/colorbars/correlation_cbar.pdf}\\
    \caption{Supervoxel-wise analysis of selected axial slices of density images for the (a) female cohort and (b) male cohort. Each image is displayed as a reference image slice, correlation maps (where non-significance is shown as black), and the correlation map overlaid on the reference image. We observe that the fatty tissue (in particular around the coronary vessels) in these slices exhibits a negative association with age.}
    \label{fig:imiomics_hu}
\end{figure}

In Fig.~\ref{fig:imiomics_hu}, we show association maps between age and density. While the density in the aorta, as well as in the left atrium and ventricle, are strongly correlated with age, we can not discard the possibility that it is an artifact of the contrast patterns. The fatty tissue around the coronary vessels exhibits a significant negative correlation with age.

The thin segment of fatty tissue on one side of the aorta exhibits a significant positive correlation to age, in both sexes.

Additionally, we observe that the bone marrow in the vertebrae has a negative association with age in females, and the bone marrow in the sternum has a negative relation with age in both sexes. 

Additional slices of supervoxel-wise association maps are presented in \ref{sec:appendixvoxelwise}.

\section{Discussion}

\paragraph{Strengths}

We conducted the study on 1388 subjects, which is a large enough sample size to obtain acceptable statistics for clear associations. The study used images acquired under a single standardized protocol designed to facilitate large-scale retrospective cohort studies.

We performed these analyses based on images acquired of individuals in the age span 50-65, which corresponds to the age span past menopause in most females (with substantial hormonal effects) and before the typical retirement age in Sweden (with substantial lifestyle changes), limiting these potentially confounding factors.

By performing different types of analyses of the data (pair-wise correlation of explicit measurements by segmentation between volumetric and density features as well as age, supervoxel-wise local volume analysis and its association to age) and observing that some relationship appeared in several analyses, we are able to be more confident in the validity and relevance of the found associations, than if observed through a single mode of analysis.

\paragraph{Limitations}
The analysis in this work has been applied to data acquired at a single site. Extending this cohort study to a multi-site setting, while having the potential to increase the statistical power by adding more data, could also introduce additional challenges due to slight variations in the application of the imaging protocol, in particular for the inter-subject deformable image registration process. Furthermore, the administration of contrast media depended on the weight of the individual which correlates with volume, and indirectly, with age. The measured density values are affected by this, and the distribution of high contrast in the organ differs accordingly.

The image registration relies on the segmentation of the heart and surrounding organs, adding complexity to the pipeline, additional runtime, and potential sources of errors.

The deformable registration of the heart could likely be improved further, to be more evenly deformed within uniform regions, such as the heart chambers. By achieving a more uniform deformation within the chambers, the proof-of-concept Imiomics analysis of the local voxel volume could then be improved to exhibit correlations closer to 1, which would likely benefit the accuracy of correlation between local voxel volume and age as well. Large subregions of the coronary vessels were successfully registered for a large fraction of the subjects (seen from visual inspection from mean images in Fig.~\ref{fig:heartregistrationexample}, and in appendix \ref{sec:appendixregistrations}), but increasing the performance of this aspect of the registration would be beneficial for any cohort-wide analysis of features related to those vessels, such as plaques and narrowing of the vessels. Another source of potential error in the registration comes from the accuracy of the segmentation. In the absence of perfectly accurate (and in particular consistent) sub-structure segmentation, the registration is likely impacted, which would affect the local volume measurements.

Even with the attempts to select an average template by considering both age and volumetric features, some bias may remain due to idiosyncratic aspects of any one subject (and corresponding volumes and shapes). A fruitful avenue of future work would include performing the analysis with average volume (generated) templates \citep{pilia2019average}, that attempt to remove more bias from the analysis than what is possible with a single template selection.

\paragraph{Future Work}
This work has focused on the development of a method for the study of the association between an individual's chronological age and cardiac morphology from CCTA image volumes. The developed framework could without modification be used to relate morphology to biomarkers attempting to model an individual's biological age, a topic of much interest and ongoing research. The inter-subject image registration methodology developed may also be used to conduct cohort saliency analysis, where saliency maps from deep learning predictions (e.g. deep age regression) are mapped to a common reference space, and their statistics are analyzed voxel-wise to provide insights into the locality of the features of importance for the black-box predictions \citep{langner2019identifying}.

Additionally, supervoxel-wise association studies relating the cardiac morphology to known risk factors and outcomes (in particular CVD and heart failure) is a future application of the developed methodology with strong potential to generate new clinical insights.

\section{Conclusion}

In this work we have developed a new method for supervoxel-wise analysis of non-imaging variables (with a focus on subject age) and CCTAs, and evaluated its performance by measuring the performance of the image registration, pair-wise correlations between explicit measurements made using image segmentation, and supervoxel-wise analysis of the local volume and explicitly measured volume variables (where we expected and observed a very high positive correlation). The method was then applied to conduct a sex-stratified association study between age and both local volume as well as density on a subset of the SCAPIS study. We observed different patterns of both age/local volume correlations and age/density correlations in males and females, indicating several differences between the sexes in how the heart's morphology changes with age. The supervoxel-wise analysis was able to find localized associations with age outside of the commonly segmented and analyzed sub-regions, showcasing one of the major advantages of supervoxel-wise studies for exploratory studies.

\section*{Acknowledgements}

This research was funded by the Swedish Heart and Lung Foundation through grant number 2022012924. 

The main funding body of The Swedish CArdioPulmonary bioImage Study (SCAPIS) is the Swedish Heart-Lung Foundation. The study is also funded by the Knut and Alice Wallenberg Foundation, the Swedish Research Council and VINNOVA (Sweden’s Innovation agency) the University of Gothenburg and Sahlgrenska University Hospital, Karolinska Institutet and Stockholm county council, Link\"{o}ping University and University Hospital, Lund University and Sk\aa ne University Hospital, Ume\aa~University and University Hospital, Uppsala University and University Hospital.

\appendix

\section{Additional Supervoxel-wise Association Maps}
\label{sec:appendixvoxelwise}

\begin{figure}[ht]
    \centering
    \textbf{Supervoxel-wise associations (female): local volume (JD) and age} \\
    \includegraphics[width=\textwidth]{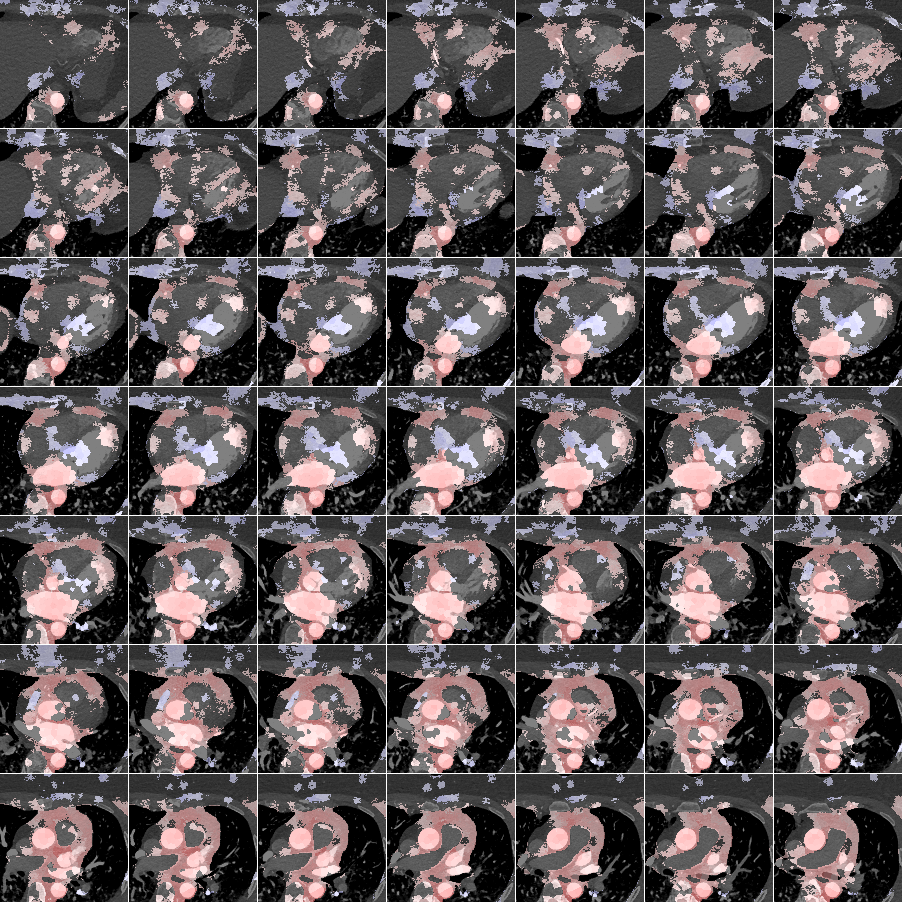}\\
    \hfill\includegraphics[width=0.3\textwidth]{figs/colorbars/correlation_cbar.pdf}
    \caption{Supervoxel-wise Pearson correlation for 49 evenly spaced axial slices, placed left to right, top to bottom.}
    \label{fig:imiomics_collage_female_jd}
\end{figure}

\begin{figure}[ht]
    \centering
    \textbf{Supervoxel-wise associations (male): local volume (JD) and age} \\
    \includegraphics[width=\textwidth]{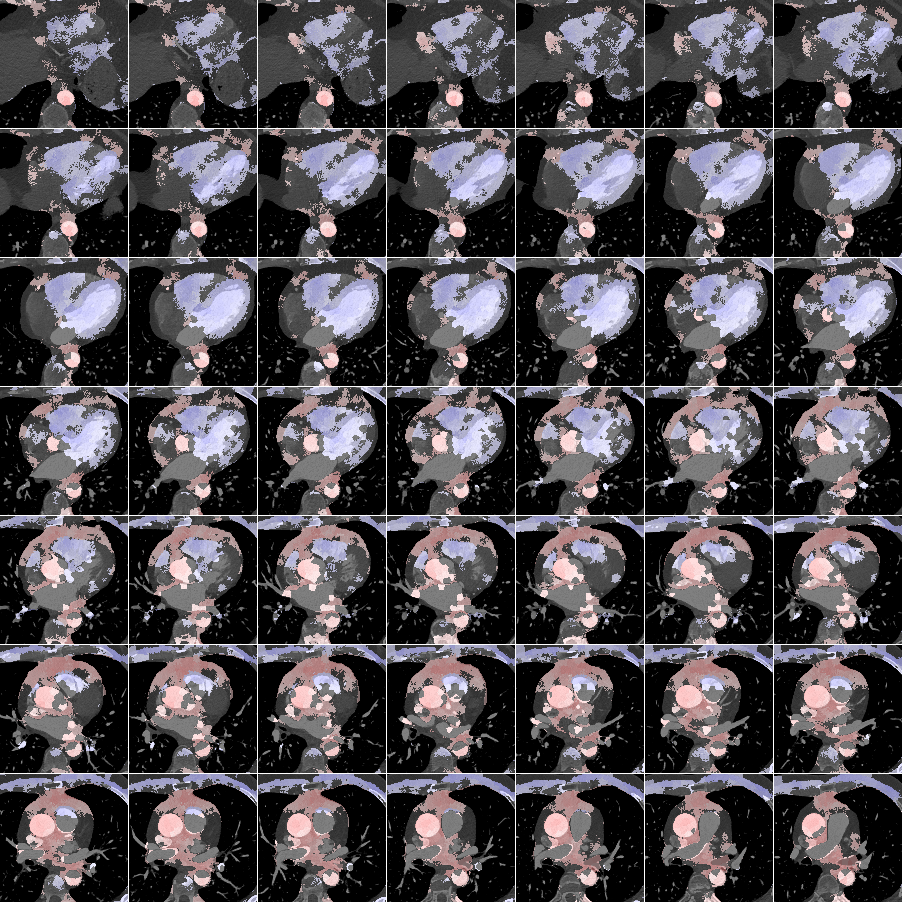}\\
    \hfill\includegraphics[width=0.3\textwidth]{figs/colorbars/correlation_cbar.pdf}
    \caption{Supervoxel-wise Pearson correlation for 49 evenly spaced axial slices, placed left to right, top to bottom.}
    \label{fig:imiomics_collage_male_jd}
\end{figure}

\begin{figure}[ht]
    \centering
    \textbf{Supervoxel-wise associations (female): density (HU) and age} \\
    \includegraphics[width=\textwidth]{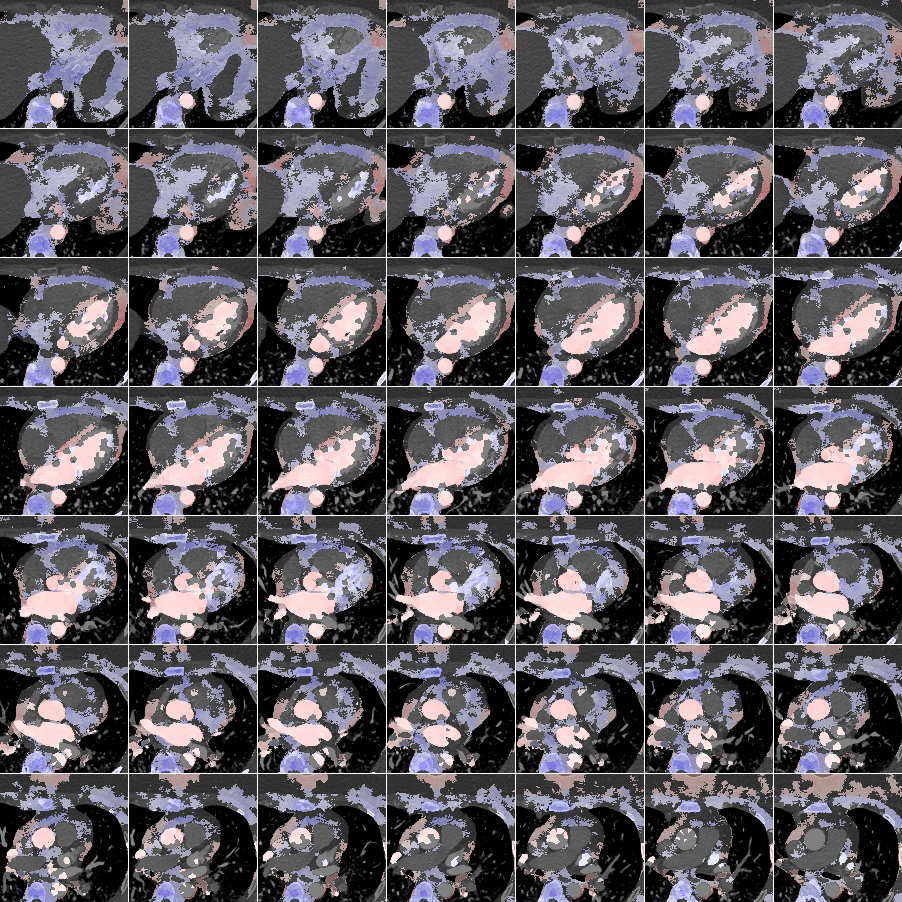}\\
    \hfill\includegraphics[width=0.3\textwidth]{figs/colorbars/correlation_cbar.pdf}
    \caption{Supervoxel-wise Pearson correlation for 49 evenly spaced axial slices, placed left to right, top to bottom.}
    \label{fig:imiomics_collage_female_hu}
\end{figure}

\begin{figure}[ht]
    \centering
    \textbf{Supervoxel-wise associations (male): density (HU) and age} \\
    \includegraphics[width=\textwidth]{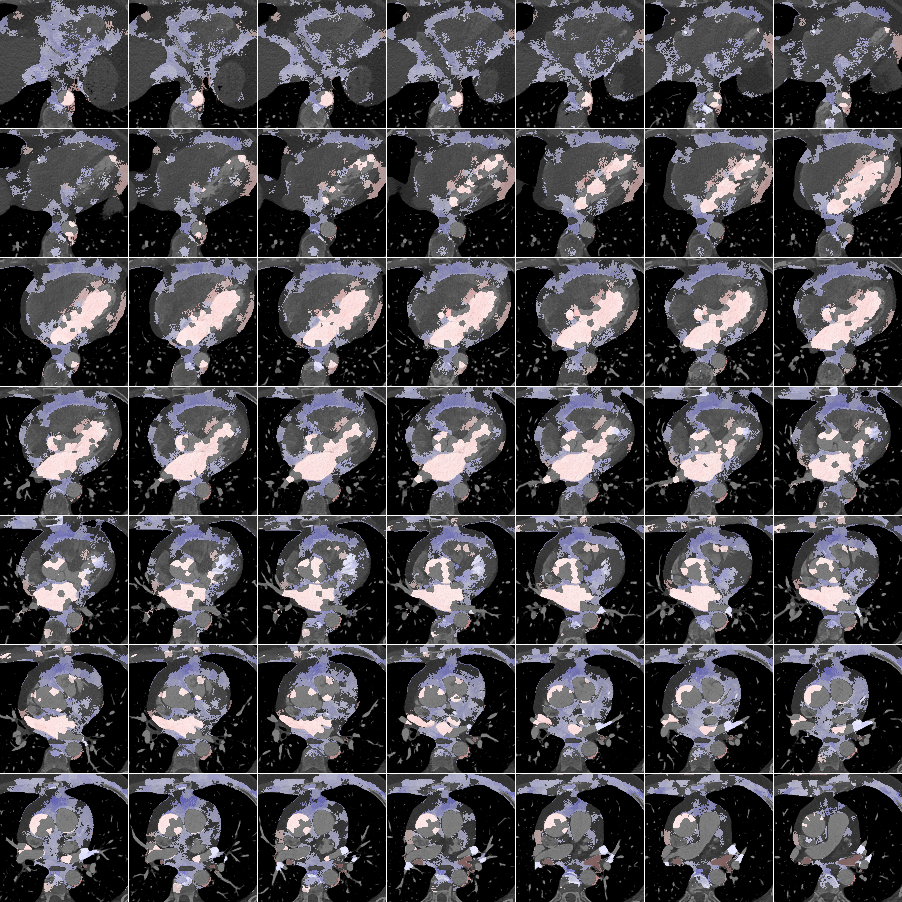}\\
    \hfill\includegraphics[width=0.3\textwidth]{figs/colorbars/correlation_cbar.pdf}
    \caption{Supervoxel-wise Pearson correlation for 49 evenly spaced axial slices, placed left to right, top to bottom.}
    \label{fig:imiomics_collage_male_hu}
\end{figure}

Here we include additional supervoxel-wise association maps for JD and age for female subjects (Fig.~\ref{fig:imiomics_collage_female_jd}, for JD and age for male subjects (Fig.~\ref{fig:imiomics_collage_male_jd}), for density and age for female subjects (Fig.~\ref{fig:imiomics_collage_female_hu}), and for density and age for male subjects (Fig.~\ref{fig:imiomics_collage_male_hu}).

\section{Additional Slices for Aggregate and Deviation Images/Jacobian Determinant Images from the Image Registration Process}
\label{sec:appendixregistrations}

\iftrue
Here we include the registration results for two additional axial slices, shown in Fig.~\ref{fig:heartregistrationappendix}.

\begin{figure}[ht]
    \centering
    \begin{turn}{90} 
        \;\;\;Female       
    \end{turn}\hfill
    \subfloat[][Reference]{\includegraphics[width=0.15\textwidth]{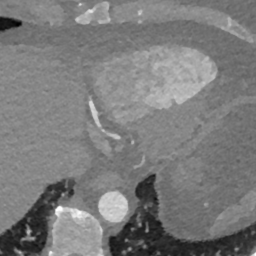}}\hfill    \subfloat[][HU Mean]{\includegraphics[width=0.15\textwidth]{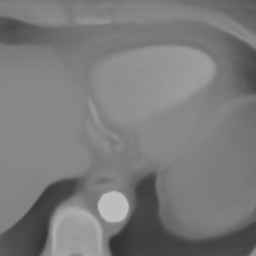}}\hfill
    \subfloat[][HU SD]{\includegraphics[width=0.15\textwidth]{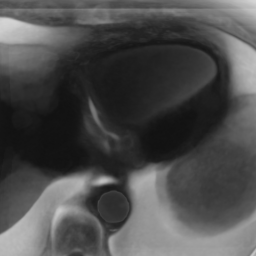}}\hfill
    \subfloat[][JD Mean]{\includegraphics[width=0.15\textwidth]{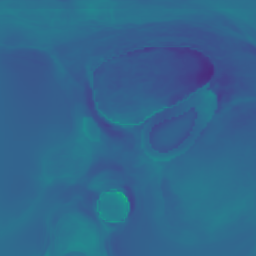}}\hfill
    \subfloat[][JD SD]{\includegraphics[width=0.15\textwidth]{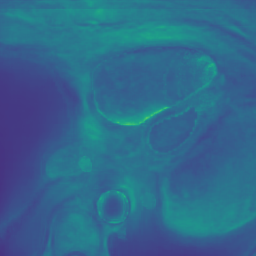}}\hfill
    \subfloat[][ICE Mean]{\includegraphics[width=0.15\textwidth]{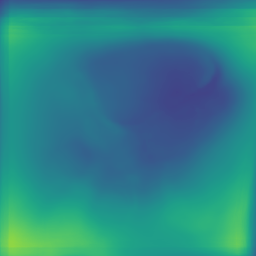}}    
    \\
    \begin{turn}{90} 
        \;\;\;\;\;Male       
    \end{turn}\hfill
    \subfloat[][Reference]{\includegraphics[width=0.15\textwidth]{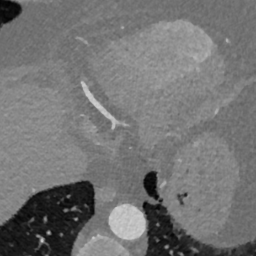}}\hfill    \subfloat[][HU Mean]{\includegraphics[width=0.15\textwidth]{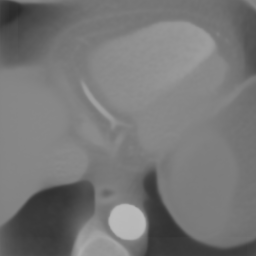}}\hfill
    \subfloat[][HU SD]{\includegraphics[width=0.15\textwidth]{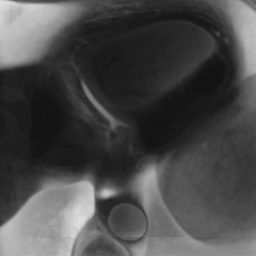}}\hfill
    \subfloat[][JD Mean]{\includegraphics[width=0.15\textwidth]{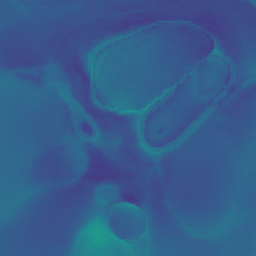}}\hfill
    \subfloat[][JD SD]{\includegraphics[width=0.15\textwidth]{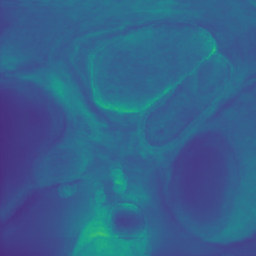}}\hfill
    \subfloat[][ICE Mean]{\includegraphics[width=0.15\textwidth]{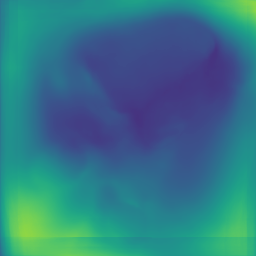}} \\
    \begin{turn}{90} 
        \;\;\;Female       
    \end{turn}\hfill
    \subfloat[][Reference]{\includegraphics[width=0.15\textwidth]{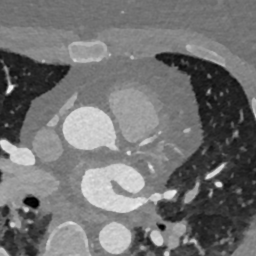}}\hfill    \subfloat[][HU Mean]{\includegraphics[width=0.15\textwidth]{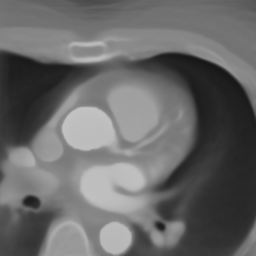}}\hfill
    \subfloat[][HU SD]{\includegraphics[width=0.15\textwidth]{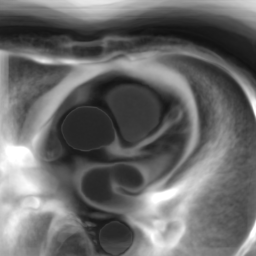}}\hfill
    \subfloat[][JD Mean]{\includegraphics[width=0.15\textwidth]{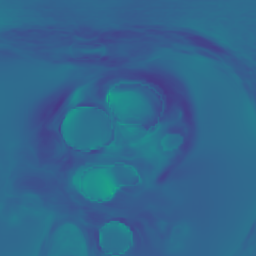}}\hfill
    \subfloat[][JD SD]{\includegraphics[width=0.15\textwidth]{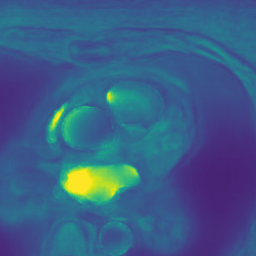}}\hfill
    \subfloat[][ICE Mean]{\includegraphics[width=0.15\textwidth]{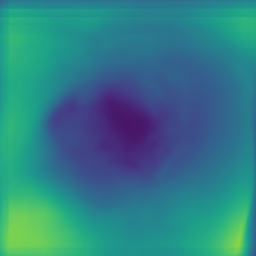}} 
    \\
    \begin{turn}{90} 
        \;\;\;\;\;Male       
    \end{turn}\hfill
    \subfloat[][Reference]{\includegraphics[width=0.15\textwidth]{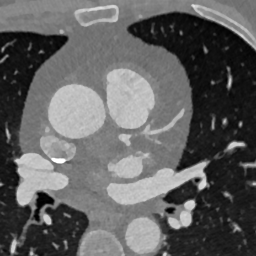}}\hfill    \subfloat[][HU Mean]{\includegraphics[width=0.15\textwidth]{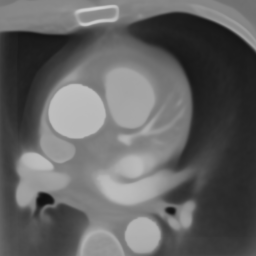}}\hfill
    \subfloat[][SHU D]{\includegraphics[width=0.15\textwidth]{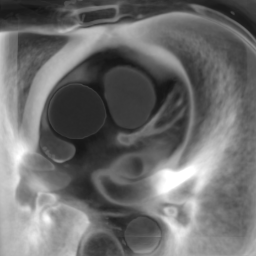}}\hfill
    \subfloat[][JD Mean]{\includegraphics[width=0.15\textwidth]{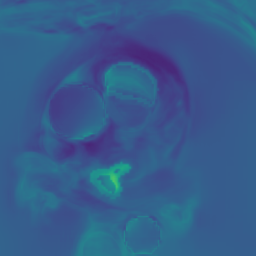}}\hfill
    \subfloat[][JD SD]{\includegraphics[width=0.15\textwidth]{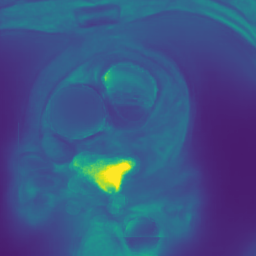}}\hfill
    \subfloat[][ICE Mean]{\includegraphics[width=0.15\textwidth]{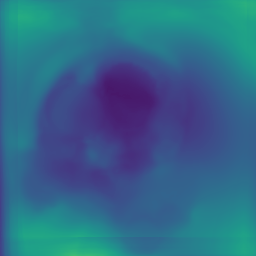}}\\
    \begin{turn}{90}
        CB       
    \end{turn}\hfill
    \includegraphics[width=0.15\textwidth]{figs/colorbars/ct_cbar.pdf}\hfill
    \includegraphics[width=0.15\textwidth]{figs/colorbars/ct_cbar.pdf}\hfill
    \includegraphics[width=0.15\textwidth]{figs/colorbars/ct_sd_cbar.pdf}\hfill
    \includegraphics[width=0.15\textwidth]{figs/colorbars/jd_cbar.pdf}\hfill
    \includegraphics[width=0.15\textwidth]{figs/colorbars/jd_sd_cbar.pdf}\hfill
    \includegraphics[width=0.15\textwidth]{figs/colorbars/ice_cbar.pdf}    
    \caption{Additional visualizations of the registration performance through example axial slices.}
    \label{fig:heartregistrationappendix}
\end{figure}
\fi

\bibliographystyle{elsarticle-harv} 
\bibliography{references}





\end{document}